\documentclass[a4paper,12pt]{article}
\usepackage[T1]{fontenc}
\usepackage[margin=0.91in]{geometry}
\usepackage{authblk}
\usepackage{breakcites}
\usepackage[utf8]{inputenc}
\usepackage{multirow}
\usepackage{multicol}
\usepackage{array,ragged2e}
\usepackage[toc,page,title]{appendix}
\usepackage{bbm,bm}
\usepackage{dsfont}
\usepackage{etoolbox}
\usepackage{url}
\usepackage{color}
\usepackage{adjustbox}
\usepackage{setspace}
\makeatletter
\patchcmd{\@makecaption}
{\parbox}
{\advance\@tempdima-\fontdimen2} 
{}{}
\makeatother 
\usepackage{graphicx}
\usepackage{subfig} 
\usepackage{enumerate}
\usepackage{comment}
\usepackage{natbib, float}
\usepackage{amsmath,amssymb,mathtools,nccmath}
\usepackage{nicefrac}
\usepackage[bottom]{footmisc}
\usepackage[colorlinks,citecolor=blue,urlcolor=blue,linkcolor=blue,filecolor=blue,breaklinks=true,backref=page]{hyperref}
\usepackage{footnotebackref}
\newcommand{\argmax}{{\operatorname{arg}\operatorname{max}}}
\def\given{\,|\,}
\makeatletter
\newcommand\notsotiny{\@setfontsize\notsotiny\@vipt\@viipt}
\makeatother
\newcommand{\indicator}[1]{\mathds{1}\negthinspace{\left( {#1} \right) }}
\newcommand{\GG}[1]{}

\makeatletter
\def\footnoterule{\kern-3\p@
	\hrule \@width 2in \kern 2.6\p@} 
\makeatother
\providecommand{\keywords}[1]{\footnotesize{\textbf{Keywords:} #1}}
\usepackage[framemethod=TikZ]{mdframed}
\usetikzlibrary{shadows}
\usepackage{environ}
\usepackage{varwidth}
\usepackage{ctable}

\newlength{\MyMdframedWidthTweak}%
\NewEnviron{reminder}[1][]{%
	\setlength{\MyMdframedWidthTweak}{\dimexpr%
		+\mdflength{innerleftmargin}
		+\mdflength{innerrightmargin}
		+\mdflength{leftmargin}
		+\mdflength{rightmargin}
	}%
	\savebox0{%
		\begin{varwidth}{\dimexpr0.7875\linewidth-\MyMdframedWidthTweak\relax}%
			\BODY
		\end{varwidth}%
	}%
	\begin{mdframed}[
		backgroundcolor=gray!20, 
		shadow=true, 
		align=center,
		shadowsize=4pt,
		roundcorner=5pt,
		userdefinedwidth=\dimexpr\wd0+\MyMdframedWidthTweak\relax, 
		#1]
		\usebox0
	\end{mdframed}
}

\title{\textbf{Clustering Longitudinal Life-Course Sequences}\\ \textbf{Using Mixtures of Exponential-Distance Models}}
\author{Keefe Murphy\textsuperscript{1},\: T. Brendan Murphy\textsuperscript{2,3},\\Raffaella Piccarreta\textsuperscript{4},\: I. Claire Gormley\textsuperscript{2,3}}
\affil[]{\small{\textsuperscript{1}~Department of Mathematics and Statistics, Maynooth University, Ireland\vspace{0.5em}}\\
	\small{\textsuperscript{2}~School of Mathematics and Statistics, University College Dublin, Ireland\vspace{0.5em}}\\
	\small{\textsuperscript{3}~Insight Centre for Data Analytics, University College Dublin, Ireland\vspace{0.5em}}\\
	\small{\textsuperscript{4}~Department of Decision Sciences, Universit\`{a} Bocconi, Milano, Italy\vspace{1em}}\\
	E-mail: \href{mailto:keefe.murphy@mu.ie}{keefe.murphy@mu.ie}}

\date{}

\begin{document}

\begin{center}
\begin{reminder}
	{\centering \color{red} \textbf{\textsf{\small This is a preprint. The revised version of this paper is published as}}\\}
	\vspace{3.5pt}
	\justifying
	{\footnotesize \noindent K. Murphy, T.\,B. Murphy, R. Piccarreta, and I.\,C. Gormley (2021) %
	{``Clustering\linebreak longitudinal life-course sequences using mixtures of exponential-\linebreak distance models''.} %
	\textit{Journal of the Royal Statistical Society: Series A\linebreak (Statistics in Society)}, 184(4): 1414--1451.
	\hfill [\href{https://doi.org/10.1111/rssa.12712}{\sf doi: 10.1111/rssa.12712}].}
\end{reminder}
\end{center}

{\let\newpage\relax\maketitle}
	
\begin{abstract}
	Sequence analysis is an increasingly popular approach for analysing life courses represented by ordered collections of activities experienced by subjects over time. Here, we analyse a survey data set containing information on the career trajectories of a cohort of Northern Irish youths tracked between the ages of 16 and 22. We propose a novel, model-based clustering approach suited to the analysis of such data from a holistic perspective, with the aims of estimating the number of typical career trajectories, identifying the relevant features of these patterns, and assessing the extent to which such patterns are shaped by background characteristics.
		
	Several criteria exist for measuring pairwise dissimilarities among categorical sequences. Typically, dissimilarity matrices are employed as input to heuristic clustering algorithms. The family of methods we develop instead clusters sequences directly using mixtures of exponential-distance models. Basing the models on weighted variants of the Hamming distance metric permits closed-form expressions for parameter estimation. Simultaneously allowing the component membership probabilities to depend on fixed covariates and accommodating sampling weights in the clustering process yields new insights on the Northern Irish data. In particular, we find that school examination performance is the single most important predictor of cluster membership.\bigskip
	
	\noindent \keywords{exponential-distance models, gating covariates, life-course sequences, model-based clustering, survey sampling weights, weighted Hamming distance.}
\end{abstract}
	
\section[Introduction]{Introduction}
\label{Section:Intro}
\normalsize

Sequence analysis (SA) is an umbrella term for tools defined to explore and describe categorical life-course data. Specifically, attention is focused on the ordered sequence of states (or activities) experienced by individuals over a given time-span (usually at $T$ equally spaced discrete time periods). Here we focus on the transition from school to work for a cohort of Northern Irish youths, using survey data obtained from the 1999 sweep of the Status Zero Survey \citep{McVicar2000, mvad2002} --- henceforth referred to as the MVAD data --- in which, for each individual, a sequence of monthly labour market activities experienced between the ages of 16 and 22 is recorded.

Typically, the goal of sequence analysis is to identify the most relevant patterns in the data. To this end, pairwise dissimilarities among sequences in their entirety are first assessed. Dissimilarity matrices are then employed to identify the most typical trajectories using, in the vast majority of applications, cluster analysis. These problems are receiving increasing attention in the demographic and social literature, also due to the increasing number of retrospective as well as prospective longitudinal studies, such as the British Household Panel Survey (BHPS)\protect\footnote{\href{https://www.iser.essex.ac.uk/bhps}{\texttt{\,https://www.iser.essex.ac.ul/bhps}}.} and  the subsequent larger and more wide-ranging UK Household Longitudinal Study (Understanding Society)\protect\footnote{\href{https://www.understandingsociety.ac.uk/}{\texttt{\,https://www.understandingsociety.ac.uk}}.}, or the Socio-Economic Panel for Germany (SOEP)\protect\footnote{\href{https://www.diw.de/en/soep}{\texttt{\,https://www.diw.de/en/soep}}.}, the National Longitudinal Surveys for the USA (NLS)\protect\footnote{\href{https://www.bls.gov/nls/}{\texttt{\,https://www.bls.gov/nls/}}.}, and the Generations \& Gender Programme for selected European countries (GGP)\protect\footnote{\href{https://www.ggp-i.org/}{\texttt{\,https://www.ggp-i.org}}.}. All of these surveys, much like the MVAD data considered in this paper, collect information about labour market activities, as well as other significant life events.

Quantifying the distance between such categorical sequences is not a trivial task. \mbox{Optimal} matching (OM), developed by \citet{Abbott1986} and extended to sociology by \citet{Abbott1990}, is popular among the SA community. OM is derived from the edit distance originally proposed in the field of information theory and computer \mbox{science} by \citet{Levenshstein1966}. The OM metric assigns costs to the different types of edits, namely insertion, deletion, and substitution. Typically, insertion and deletion are assigned a cost~of $1$ while substitution costs are allowed to vary. However, specifying these costs often~\mbox{involves} subjective choices, which may lead to violations of the triangle inequality if not done~\mbox{carefully}. Several proposals in the literature introduced criteria to improve or guide the choice of costs in OM; \citet{MunozBullon2003}, for instance, estimate the substitution-cost matrix in a data-driven fashion using the between-states transition rates. \mbox{Alternative~dissimilarity} criteria have also been introduced to allow control over the importance assigned to the characteristics of the sequences (namely, the collection of experienced states, their \mbox{timing}, or their duration) in the assessment of their differences: see \citet{Studer2016}~for~an excellent discussion. Even so, there are no results proving that one procedure is~\mbox{superior~to} others and the choice of dissimilarity measure remains a fundamental question for~\mbox{researchers}.\enlargethispage{\baselineskip}

Given a dissimilarity matrix $\mathbf{D}$ obtained from a set of sequences $\mathbf{S}=\left(\mathbf{s}_1,\ldots,\mathbf{s}_n\right)$,~where~$n$ is the number of subjects, cluster analysis is usually applied to group sequences and~\mbox{identify} the most typical trajectories experienced by the sampled individuals. Heuristic~\mbox{clustering} algorithms, either hierarchical or partitional, are typically used. In many applications, it is also of interest to relate sequences to a set of baseline covariates. Within the~\mbox{described} framework, this is solely done by relating the uncovered hard clustering partition to covariates using, for example, multinomial logistic regression (MLR). This approach was adopted in \citet{mvad2002}, after applying Ward's agglomerative~\mbox{clustering} algorithm \citep{Ward1963} to an OM dissimilarity matrix to obtain $G=5$ clusters of the MVAD sequences, without performing model selection. Such an approach~is~\mbox{questionable~from} a few points of view. Firstly, the original sequences are substituted by a~categorical~\mbox{variable} indicating cluster membership, thus disregarding the heterogeneity within clusters. This is clearly only sensible when the clusters are sufficiently homogeneous\, otherwise sequences which are weakly related to clusters would be regarded as similar to those in their cluster. However, a clear clustering structure can often be obtained only by increasing the number of clusters (often with some clusters possibly small in size). More importantly, suitable partitions do not necessarily lead to suitable response variables as input for the MLR. It thus seems desirable to cluster sequences and relate the clusters to the covariates simultaneously.

Thus, the aim of our analysis is three-fold; to estimate the number of typical trajectories in the MVAD data, to identify the relevant features of these patterns, and to establish to what extent such patterns are shaped by the individuals' background characteristics, as~captured by a set of baseline covariates measured at age 16. To address these issues, we propose to cluster the MVAD sequences in a model-based fashion, allowing the covariates~to affect~the soft cluster membership probabilities, rather than leaving them exogenous to the~\mbox{clustering} model. This permits us to better understand if and to what extent the typical sequence patterns characterising each cluster are affected by specific covariates. Model-based~\mbox{clustering} methods typically assume that the data arise from a finite mixture of $G$ distributions; \citet{Bouveyron2019} provide an excellent overview. In principle, any distribution(s) can be used, though the term `model-based clustering' was popularised by \citet{Banfield1993}, in which the component distributions are assumed to be parsimoniously parameterised multivariate Gaussians with component-specific parameters. Such models have been recently extended to the mixture of experts setting \citep{GormleySchnatter2019} to facilitate dependence on fixed covariates \citep{Murphy2020}. However, these models can be problematic when applied to dissimilarity~\mbox{matrices}, either due to non-identifiability or because the input data are usually far from Gaussian. This problem cannot be addressed by applying multidimensional scaling to $\mathbf{D}$ because the resulting low-dimensional configuration is also typically far from Gaussian. Notably, our attempts to fit non-Gaussian mixtures in these settings did not yield useful results.

Another popular framework for clustering categorical data is latent class analysis (LCA; \citealt{Lazarsfeld1968})). \citet{Agresti2002} shows the connection between model-based clustering and LCA. Such models are finite mixtures in which the component distributions are assumed to be multi-way cross-classification tables with all variables mutually independent. Latent class regression models (LCR; \citealt{Dayton1988}) are particularly interesting, because their connection to the mixture of experts framework  permits the inclusion of covariates to predict the latent class memberships. However, fitting such models is challenging when the sequence length, the number of categories, or the number of latent classes are even moderately large, due to the explosion in the number of parameters.

Evidently, there is a conflict of perspectives between the model-based and the~\mbox{heuristic}, distance-based approaches to clustering in the SA community. For this reason, and the~\mbox{others} mentioned above, we model the sequences directly (in the sense that the sequences themselves are treated as inputs, rather than $\mathbf{D}$) with the implicit substitution costs which define the distance metric being estimable parameters of a generative probability model rather~than inputs (either estimated or subjectively specified), via $\mathbf{D}$, to a heuristic clustering algorithm. This is achieved using parsimonious mixtures of exponential-distance models, which~typically depend on a central sequence and a precision parameter in a way that relates~to~the chosen distance metric. Our framework for analysing the MVAD data, as a model-based approach which nonetheless relies on distances, thus reconciles the aforementioned conflict.\enlargethispage{\baselineskip}

Mostly for reasons of computational convenience, we use dissimilarities based on simple matching, in particular the Hamming distance \citep{Hamming1950}. Although the focus on substitution operations has the sociological advantage of targeting trajectories with contemporaneous similarities --- in contrast to the prohibited insertion and deletion operations, which focus on matching states irrespective of their timing --- this distance is liable to suffer from temporal rigidity, since anticipations and/or postponements of the same choices in life courses are not accounted for. Hence, similar sequences shifted by one time period may be maximally distant from one another. While misalignment is less of a concern for sequences exhibiting long durations in the same state, we address the issue using weighted variants of the Hamming distance, characterised by a range of constraints on the precision parameters in the mixture setting. This leads to the novel MEDseq model family, which can be seen as similar to a version of the $k$-medoids/PAM algorithm \citep[Chapter 2]{Kaufman1990} based on the Hamming distance with some restrictions relaxed. We defer the comparison to Section \ref{Section:INIT} as the parallels relate to technical issues of model estimation.

Importantly, information is also available with the MVAD data on the survey sampling weights, which are only incorporated in the MLR stage of the analysis in \citet{mvad2002}. While sampling weights can be incorporated into heuristic clustering algorithms, such as Ward's hierarchical clustering (by weighting the linkages between clusters) or $k$-medoids, and subsequently in the MLR, one of the advantages of our approach is that both the covariates and the weights are incorporated simultaneously. This is achieved by leveraging the model-based paradigm; the weights are incorporated by appropriately weighting the likelihood function and the covariates are incorporated by assuming they influence the soft component membership probabilities.

MEDseq models, like standard SA heuristic clustering algorithms and LCA models, approach the clustering task from the holistic perspective of treating trajectories as \emph{whole} units of analysis, in order to uncover groups of similar sequences. In contrast, a number of multistate models employing finite mixtures with Markov components (e.g. \citealt{Melnykov2016, Pamminger2010}) or with hidden Markov components \citep{Helske2016} have recently attained popularity for the analysis of categorical sequence data. Such models focus on modelling instantaneous transitions within the life course and on factors that might explain the probability of experiencing them. As described by \citet{Wu2000}, this amounts to a difference between considering sequences in their entirety under the MEDseq framework or as time-to-event processes under the Markovian framework. Indeed, as our aim is to establish sequence typologies for the MVAD data, a holistic approach is preferable to Markovian approaches. The former concentrates on questions of global similarities and considers the full richness of the trajectories without discarding the details of episode ordering, duration, or transition \citep{MunozBullon2003}, while the latter framework makes the often unsuitable simplifying assumption that trajectories can be efficiently summarised only by their recent past \citep{Piccarreta2019}.

The remainder of the article is organised as follows. Section \ref{Section:MVADdata} presents some exploratory analysis of the MVAD data. 
Section \ref{Section:Modelling} develops the MEDseq family of~\mbox{mixtures of} exponential-distance models. Section \ref{Section:Estimation} describes the model fitting procedure and discusses factors affecting performance. Section \ref{Section:Results} presents results for the MVAD data, including appli\-cations of MEDseq models and comparisons to other methods. The insights gleaned~from the MVAD data under the optimal MEDseq model are summarised in Section \ref{Section:Discussion}. \mbox{The~paper} concludes with a discussion on the MEDseq methodology and potential future extensions~in Section \ref{Section:Conclusion}. A software implementation of the full MEDseq model family is provided by~the associated \textsf{R} package \texttt{MEDseq} \citep{MEDseqR2021}. The package was developed specifically for this application and is available from \href{https://www.r-project.org}{\texttt{https://www.r-project.org}} \citep{R2021}.

\section[Status Zero Survey: MVAD Data]{Status Zero Survey: MVAD Data}
\label{Section:MVADdata}

The term `MVAD data' refers throughout to a cohort of $n=712$ Northern Irish youths aged 16 and eligible to leave compulsory education as of July 1993 who were observed at monthly intervals until June 1999 as part of the Status Zero Survey \citep{Armstrong1997,McVicar2000,mvad2002}. The subjects were interviewed about the labour market activities they experienced, distinguishing between employment (EM), further education (FE), higher education (HE), joblessness (JL), school (SC), or training (TR). Each observation $i$ is represented by an ordered categorical sequence of length $T=72$,~with an alphabet $\bm{\mathcal{V}}$ of size $v=6$ possible states, e.g. $\mathbf{s}_i=\left(s_{i,1},s_{i,2},\ldots,s_{i,72}\right)^\top\negthinspace= \left(\mbox{SC,SC,}\ldots\mbox{,TR,TR,}\ldots\mbox{,EM,EM}\right)^\top$. The sequences share a common length, the time periods are equally spaced, and there are no missing data. 

In the context of the Northern Irish education system at the time, SC refers to secondary school, which may be a grammar~school to which entrance is granted upon completion of an exam. At age 16, students take General Certificate of Secondary Education (GCSE) examinations; students who do well are eligible to continue in school for a further two years (to sit A-level exams) or to leave, e.g. to~a training/apprenticeship programme (TR). Further education (FE) is distinguished from higher education (HE); FE typically refers to applied post-GCSE courses while HE refers to third-level/university courses, typically pursued at age 18 after the successful completion of A-level exams. Notably, the transitions $\mbox{HE}\rightsquigarrow\mbox{SC}$ and $\mbox{TR}\rightsquigarrow\mbox{HE}$ are never observed. 

It is of interest to relate the MVAD sequences to covariates in order to understand whether different characteristics (related to gender, community, geographic and social conditions, and personal abilities) impact on the school-to-work trajectories. These covariates are summarised in Table \ref{Table:MVADcovars}. All covariates were measured at the age of 16 (i.e. at the start of the study period in July 1993), with the exception of `Funemp' and `Livboth', and are thus static background characteristics. As achieving 5 or more grades at A--C in GCSE exams is the traditional cut-off point for progression to the additional two-years of secondary school required for a transition to HE, we expect the `GCSE5eq' covariate in particular to be strongly associated with the clustering. 

\begin{figure}[H]
	\centering
	\begin{minipage}[tc]{0.49\textwidth}
		\centering
		\includegraphics[width=1.025\linewidth,keepaspectratio]{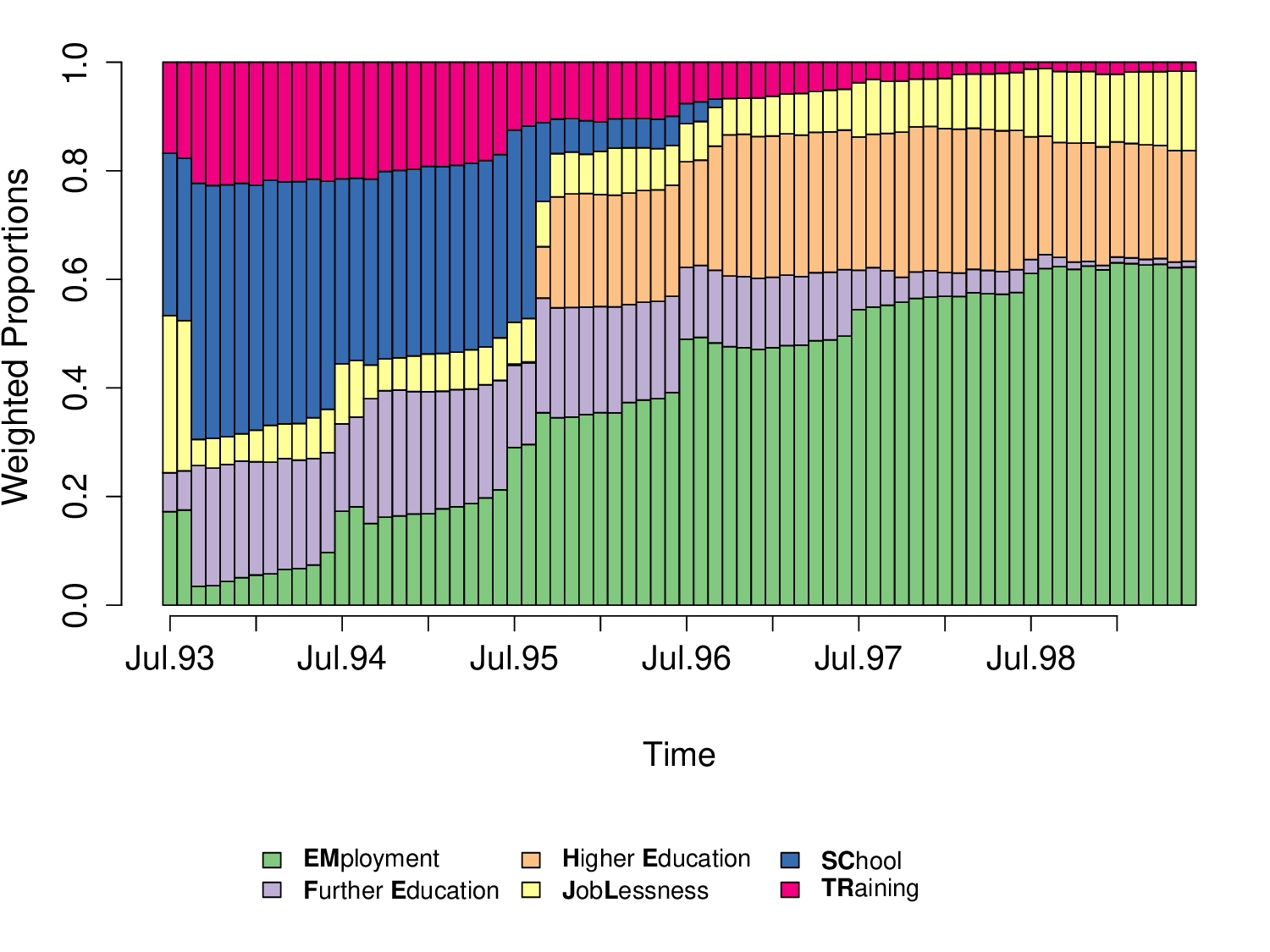}
		\caption{Overall state distribution for the weighted MVAD data, coloured by state. \label{Plot:MVADstatedist}}
	\end{minipage}\hfill%
	\begin{minipage}[tc]{0.49\textwidth}
		\centering
		\includegraphics[width=1.025\linewidth,keepaspectratio]{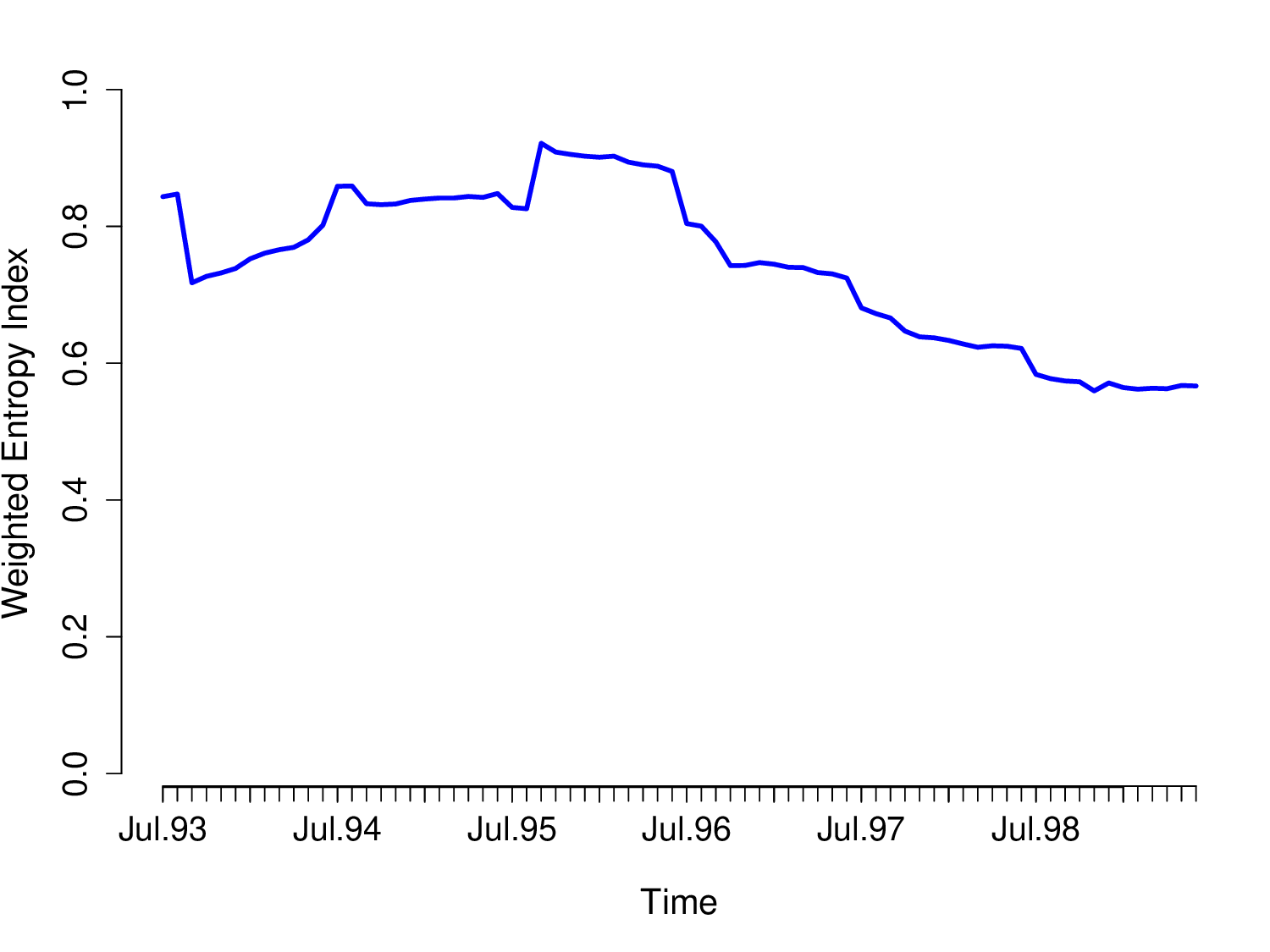}
		\caption{Transversal entropy plot for the weighted MVAD data.\label{Plot:MVADEntropy}}
	\end{minipage}
\end{figure}

The MVAD data also come with associated observation-specific survey sampling weights, which depend on the `Grammar' and `Location' covariates. Specifically, the sample was stratified in such a way that a predetermined number of subjects were in each state, for each location and both school types, immediately after the end of the compulsory education period \citep{Armstrong1997}.
\begin{table}[H]
	\caption{Available covariates for the MVAD data set. For binary covariates, the event denoted by $1$ is indicated. Otherwise, the levels of the categorical covariate `Location' are grouped in curly brackets.\label{Table:MVADcovars}}
	\scriptsize
	\extrarowheight 2.5pt
	\centering
	\begin{tabular}[pos=center]{l l}
		\specialrule{.1em}{.01em}{.01em} 
		Covariate & Description\\
		\hline\hline
		Catholic & $1$=yes\\
		FMPR & SOC code of father's current or most recent job as of the beginning of the survey,\\ & $1$=SOC1 (Standard Occupational Classification: professional, managerial, or related)\\
		Funemp & Father's employment status as of June 1999, $1$=employed\\
		GCSE5eq & Qualifications gained by the end of compulsory education,  $1$=5+ GCSE grades at A--C, or equivalent\\
		Gender & $1$=male\\
		Grammar & Type of secondary education, $1$=grammar school\\
		Livboth & Living arrangements as of June 1995, $1$=living with both parents\\
		Location & Location of school, one of five Education and Library Board areas in Northern Ireland,\\
		& \{Belfast, North Eastern, South Eastern, Southern, Western\}\\
		\specialrule{.1em}{.01em}{.01em} 
	\end{tabular}
\end{table}
\clearpage
The MVAD data are available in the \textsf{R} packages \texttt{MEDseq} and \texttt{TraMineR} \citep{TraMineR2011}. As the data have been used to illustrate some of the functionalities of the \texttt{TraMineR} package in its associated vignette\protect\footnote{\href{https://cran.r-project.org/web/packages/TraMineR/vignettes/TraMineR-state-sequence.pdf}{\texttt{\,https://cran.r-project.org/web/packages/TraMineR/vignettes/TraMineR-state-sequence.pdf}}.}, interesting features of an exploratory analysis of the data can be found therein. However, we reproduce plots of the transversal state distributions in Figure \ref{Plot:MVADstatedist} and the transversal entropies in Figure \ref{Plot:MVADEntropy}, i.e. the Shannon entropies of the state distributions at each time point \citep{Billari2001}, with the sampling weights accounted for in both cases. Notably, fewer than $v$ states are observed in certain months.

Figure \ref{Plot:MVADstatedist} shows that the number of subjects who found employment increased over time. Conversely, fewer students were in training or further education by the end of the observation period. Most students appear to have entirely left school within 2/3 years of the commencement of the survey. Finally, while students only reached the age of 18 and began to pursue higher education from July 1995 onwards, a number of students had already pursued further education during the two preceding years. Figure \ref{Plot:MVADEntropy} confirms that the level of heterogeneity in the state distribution varies over time. In particular, the entropy declines after Sep 1995, by which point most students had left school.
Interestingly, many students were jobless during the first two months of observation. As the vast majority of cases notably remained in the same state in this period, which coincided with the summer break from school, all subsequent analyses are conducted on a version of the data with the first two time points removed. Hence, we work hereafter with sequences of length $T=70$, commencing with the return to school in September 1993. As the sampling design depends on `Grammar' and `Location', the term `all covariates' henceforth refers to all \emph{other} covariates in Table \ref{Table:MVADcovars}. While \citet{Murphy2020} show that the same covariate can affect more than one part of a mixture of experts model, and in different ways, removing the quantities used to define the weights eases the interpretability of the results.

\section[Modelling]{Modelling}
\label{Section:Modelling}

In this section, we introduce the novel family of MEDseq models. The exponential-distance model is described in Section \ref{Section:EDmodels}, extended to account for the sampling weights in \mbox{Section} \ref{Section:Weights}, expanded into a family of mixtures in Section \ref{Section:MEDseq}, and finally embedded within the~mixture of experts framework in Section \ref{Section:Gating} in order to accommodate the available covariates.

\subsection[Exponential-Distance Models]{Exponential-Distance Models}
\label{Section:EDmodels}
\enlargethispage{\baselineskip}

For an arbitrary distance metric $\smash{\mathrm{d}\negthinspace\left(\cdot,\cdot\right)}$, location parameter $\smash{\boldsymbol{\theta}}$, precision parameter $\lambda$, and set of all possible sequences $\smash{\bm{\mathcal{S}}_v^T}$, the probability mass function (PMF) of an exponential-distance model (EDM) for sequences is
\begin{equation}f\negthinspace\left(\mathbf{s}_i\given \boldsymbol{\theta}, \lambda, \mathrm{d}\right) = \frac{\exp\negthinspace\left(-\lambda\mathrm{d}\negthinspace\left(\mathbf{s}_i, \boldsymbol{\theta}\right)\right)}{\sum_{\boldsymbol{\sigma}\,\in\,\bm{\mathcal{S}}_v^T}\exp\negthinspace\left(-\lambda\mathrm{d}\negthinspace\left(\boldsymbol{\sigma}, \boldsymbol{\theta}\right)\right)} = \Psi\negthinspace\left(\lambda, \boldsymbol{\theta}\given T, v\right)^{-1}\exp\negthinspace\left(-\lambda\mathrm{d}\negthinspace\left(\mathbf{s}_i, \boldsymbol{\theta}\right)\right),\vspace{-0.5ex}\label{eq:EDmodel}
\end{equation}
\noindent with the corresponding log-likelihood function given by\vspace{-1ex}
\begin{equation}
\ell\negthinspace\left(\boldsymbol{\theta},\lambda\given\mathbf{S}, \mathrm{d}\right) = \sum_{i=1}^n\log f\negthinspace\left(\mathbf{s}_i\given \boldsymbol{\theta}, \lambda, \mathrm{d}\right) = -\lambda\sum_{i=1}^n\mathrm{d}\negthinspace\left(\mathbf{s}_i,\boldsymbol{\theta}\right) - n\log\Psi\negthinspace\left(\lambda,\boldsymbol{\theta}\given T, v\right).\label{eq:MEDll}\vspace{-1ex}
\end{equation}
Probabilistic EDMs, as here, in which the kernel of the distance-based mass or density~is~the inverse of the exponentiated measure of distance from a \mbox{prototype}, weighted by a~\mbox{precision} parameter, were originally formalised by \citet{Diaconis1988}. Such models are analogous~to (among others) both the Gaussian distribution and the von Mises-Fisher distribution for~data on the unit hypersphere \citep{Banerjee2005}, characterised by the squared Euclidean and cosine distances from the mean, respectively. The EDM in \eqref{eq:EDmodel} is also similar to the Mallows model for~permu\-tations \citep{Mallows1957, Fligner1986}. Notably, mixtures of Mallows models have previously been used to cluster rankings \citep{Murphy2003}.

Here, we only consider models with $\lambda \geq 0$. When $\lambda=0$, the distribution of sequences is uniform. For $\lambda > 0$, the central sequence $\smash{\boldsymbol{\theta}=\left(\theta_1,\ldots,\theta_T\right)}$~is the mode,~i.e. the sequence with highest probability, and the probability of any other sequence decays\linebreak exponentially as its distance from $\boldsymbol{\theta}$ increases. The precision parameter $\lambda$ controls the~speed of this decay. Larger $\lambda$ values cause sequences to concentrate around $\boldsymbol{\theta}$, tending toward a point-mass as $\lambda \to \infty$. Notably, $\lambda$ is not identifiable when all sequences are identical.

The log-likelihood in \eqref{eq:MEDll} is generally intractable, as the normalising constant $\Psi\negthinspace\left(\lambda,\boldsymbol{\theta}\given T, v\right)$ depends on the parameter $\lambda$ (under OM and other, more complicated distances, it can also depend on $\boldsymbol{\theta}$), as well as the fixed constants $T>1$ and $v>1$, and requires a sum over all possible sequences. With reference to the MVAD data, for example, computing $\Psi\negthinspace\left(\lambda,\boldsymbol{\theta}\given T, v\right)$ is practically infeasible as there are $\smash{\mathrm{card}\negthinspace\left(\bm{\mathcal{S}}_v^T\right)=v^T=6^{70}}$ possible sequences. Fortunately, however, the normalising constant exists in closed form under the Hamming distance, $\smash{\mathrm{d_H}\negthinspace\left(\mathbf{s}_i,\mathbf{s}_j\right) = \sum_{t=1}^T \indicator{s_{i,t} \neq s_{j,t}}}$, in a manner which facilitates direct enumeration and crucially does not depend on $\boldsymbol{\theta}$, as a sum with only $T+1$ terms. Consider, for example, the Hamming distances between all ternary ($v=3$) sequences of length $T=4$. From the arbitrary reference sequence $(0,0,0,0)$, there is $1$ count of a distance of $0$, $8$ counts of a distance of $1$, $24$ counts of a distance of $2$, $32$ counts of a distance of $3$, and $16$ counts of a distance of $4$. Thus, $\Psi_{\mathrm{H}}\negthinspace\left(\lambda\given T=4, v=3\right)=e^{0} + 8e^{-\lambda} + 24e^{-2\lambda} + 32e^{-3\lambda} + 16e^{-4\lambda}$. Hence, the normalising constant under the Hamming distance metric depends on the parameter $\lambda$, the sequence length $T$, and the number of categories $v$, and simplifies greatly:\vspace{-0.5ex}
\begin{equation}
\Psi_{\mathrm{H}}\negthinspace\left(\lambda\given T, v\right) =\sum_{p=0}^{T} {T\choose p}\left(v-1\right)^p\exp\negthinspace\left(-\lambda p\right) = \left(\left(v-1\right)e^{-\lambda} + 1\right)^T\negthinspace.\label{eq:PsiH}\vspace{-0.5ex}
\end{equation}
\indent Inspired by the generalised Mallows model \citep{Irurozki2019}, the EDM in \eqref{eq:EDmodel} based on the Hamming distance can be extended to one based on the weighted Hamming distance. By introducing $T$ precision parameters $\smash{\lambda_1,\ldots,\lambda_T}$, one for each time point (i.e. sequence position), and expressing the exponent in \eqref{eq:EDmodel} as $\smash{\mathrm{d_{WH}}(\mathbf{s}_i,\boldsymbol{\theta} \given \lambda_1,\ldots,\lambda_T)=\sum_{t=1}^T\lambda_t\indicator{s_{i,t} \neq \theta_t}}$ rather than $\smash{\lambda\mathrm{d_H}(\mathbf{s}_i,\boldsymbol{\theta})=\lambda\sum_{t=1}^T\indicator{s_{i,t} \neq \theta_t}}$, different time periods can contribute differently to the overall distance, weighted according to the period-specific precision parameters. Thus, the distance from $\boldsymbol{\theta}$ to $\mathbf{s}_i$ under $\smash{\mathrm{d_{WH}}\negthinspace\left(\cdot,\cdot\given\cdot\right)}$ becomes a sum of the $\lambda_t$ values associated~with each time point which differs from the corresponding $\theta_t$, across the whole sequence. This~acts as implicit variable selection and allows modelling situations in which there is high consensus regarding the state values of some time periods, with large uncertainty about the values~of others. Accounting for the alignment of contemporaneous matchings in this way helps to prevent sequences with the same (Hamming) distance from $\smash{\boldsymbol{\theta}}$ from having the same probability. Given that sequences equidistant from $\boldsymbol{\theta}$ can nevertheless exhibit element-wise mismatches between themselves, this may help later, in the mixture setting, to induce stronger~between-cluster separation and within-cluster homogeneity. The non-constant transversal entropies in Figure \ref{Plot:MVADEntropy} suggest that this extension may also be fruitful in terms of capturing different degrees of dispersion in the state distributions of the MVAD data over time. Crucially,~the various benefits outlined above can be achieved without any tractability sacrifices. The log-likelihood in \eqref{eq:MEDll} is merely rewritten with the weighted Hamming distance decomposed into its $T$ components and the normalising constant in \eqref{eq:PsiH} also modified accordingly:
\begin{equation*}
\ell\negthinspace\left(\boldsymbol{\theta},\lambda_1,\ldots,\lambda_T\given\mathbf{S},\mathrm{d_{WH}}\right) =
-\sum_{i=1}^n\left\lbrack\sum_{t=1}^T\Big(\lambda_t\indicator{s_{i,t} \neq \theta_t} + \log\negthinspace\big(\negthinspace\left(v-1\right)e^{-\lambda_t} + 1\big)\Big)\right\rbrack.\label{eq:MEDllCU}
\end{equation*}
\indent Though other dissimilarity measures are available for sequences, we henceforth~\mbox{consider} measures based on the Hamming distance only, chiefly for the computational reasons~\mbox{outlined} above. In our setting, $\smash{\lambda\mathrm{d_{H}}\negthinspace\left(\cdot,\cdot\right)}$ can be seen as a special case of OM with all substitution costs set to $\lambda$ and no insertions or deletions. As it has time-varying substitution costs, $\smash{\mathrm{d_{WH}}\left(\cdot,\cdot\given\cdot\right)}$ is similar to the dynamic Hamming distance \citep{Lesnard2010}, a prominent alternative to OM. However, such costs in our models are always assumed to be common with respect to each pair of states. Hence, $\smash{\mathrm{d_{WH}}\negthinspace\left(\cdot,\cdot\given\cdot\right)}$ equates to the Gower distance between nominal variables \citep{Gower1971} with equally weighted states and unequally weighted time points.
 
\subsection[Incorporating Sampling Weights]{Incorporating Sampling Weights}
\label{Section:Weights}

Sampling weights are often associated with life-course data, as the data typically arise from surveys where the weights are used to correct for representivity bias under stratified sampling designs. Following \citet{Chambers2003}, the sampling weights $\smash{\mathbf{w}=\left(w_1,\ldots,w_n\right)}$ are incorporated into the EDM by exponentiating the likelihood of each sampled unit by the attached weight $\smash{w_i}$, which is akin to unit $i$ being observed $\smash{w_i}$ times. The resultant pseudo likelihood $\smash{\mathcal{L}^{\mathbf{w}}\negthinspace\left(\cdot\given\cdot\right)}$ reweights the likelihood contribution for each unit in order to rebalance the information in the observed sample to approximate the balance of information in the target finite population. The sampling weights $\mathbf{w}$ are thus interpretable as being inversely proportional to the unit inclusion probabilities, remain fixed, and are confined to those included in the sample. Notably, $\smash{f\negthinspace\left(\mathbf{s}_i\given \boldsymbol{\theta}, \lambda, \mathrm{d}\right)^{w_i} \propto f\negthinspace\left(\mathbf{s}_i\given \boldsymbol{\theta}, w_i\lambda, \mathrm{d}\right)}$, such that the weights induce a unit-specific rescaling of the precision parameter; it follows that the observed data are independent but not identically distributed.\enlargethispage{1em}

A secondary benefit of incorporating weights is that it facilitates computational gains in the presence of duplicate cases. Such duplicates are likely when dealing with discrete life-course data. This non-uniqueness can be exploited using likelihood weights for computational efficiency, by fitting models to the subset of unique sequences only, weighted by the sum of the sampling weights (if available, otherwise $\smash{w_i=1\:\forall\:i}$) across each corresponding set of duplicates. In modifying $\mathbf{w}$ in this way, cases with different sampling weights which are otherwise duplicates are also treated as duplicates, in such a way that the (pseudo) likelihood is unaltered. The number of duplicates clearly lowers when considering both the sequences themselves and their associated covariate patterns. In particular, all cases are unique when there are continuous covariates. Nonetheless, in the MVAD data, and in many applications, the covariates are all categorical. Hence, exploiting non-uniqueness in this manner can be extremely computationally convenient. For instance, only $490$ of~the $n=712$ sequences in the MVAD data set are distinct. However, to avoid notational confusion, all subsequent expressions are written as though duplicate cases have not been discarded.

Though the weights for the MVAD data sum to $\approx711.52$, we henceforth follow \citet{Xu2013} in always assuming that the weights have been normalised to sum to the sample size $n$. In doing so, subsequent expressions are simplified further and the use of model selection criteria (see Section \ref{Section:Selection}) relying on the pseudo likelihood is facilitated. While the resultant rescaling of the MVAD weights is negligible, we note that multiplying $\mathbf{w}$ by a scalar does not affect parameter estimation. 

\subsection[A Family of Mixtures of Exponential-Distance Models]{A Family of Mixtures of Exponential-Distance Models}
\label{Section:MEDseq}

Extending the EDM based on the Hamming distance with sampling weights to the model-based clustering setting yields a pseudo likelihood function of the form
\begin{equation*}
\mathcal{L}^{\mathbf{w}}\negthinspace\left(\boldsymbol{\theta}_1,\ldots,\boldsymbol{\theta}_G,\lambda\given\mathbf{S},\mathbf{w},\mathrm{d_H}\right) = \prod_{i=1}^{n}\left\lbrack\sum_{g=1}^{G}\tau_g\frac{\exp\negthinspace\left(-\lambda\mathrm{d_H}\negthinspace\left(\mathbf{s}_i,\boldsymbol{\theta}_g\right)\right)}{\left(\left(v-1\right)e^{-\lambda} + 1\right)^T}\right\rbrack^{w_i}\negmedspace,
\end{equation*}
where the mixing proportions $\tau_1,\ldots,\tau_G$ are positive and sum to $1$. Thus, the clustering approach is both model-based and distance-based, thereby bridging the gap between these two `cultures' in the SA community.

The mixture setting naturally suggests a further extension, whereby the precision param\-eter $\lambda$ can be constrained or unconstrained across clusters, in addition to the~\mbox{aforementioned} allowance for the precision parameters to vary (or not) across time~points. Within a~\mbox{family}~of models we term `MEDseq', we thus define the \textsf{CC}, \textsf{UC}, \textsf{CU}, and \textsf{UU} models, where the~first letter denotes whether precision parameters are constrained (\textsf{C}) or unconstrained (\textsf{U}) across clusters and the second denotes the same across time points. Notably, all models~\mbox{deviate} from the simple matching distance on which they are based, as even the most~constrained~\textsf{CC} model could be said to employ a weighted variant thereof, by virtue of allowing for $\lambda \neq 1$. The model family allows moving between more parsimonious models and more~heavily~parameterised, flexible models which may provide a better fit to the data. As the precision parameters relate to the substitution costs characterising variants of the~\mbox{Hamming distance}, quantities used to define the overall distance measure are allowed to vary in different ways, while still being treated as model parameters rather than inputs. In particular,~\mbox{models~with} names beginning with \textsf{U} reflect scenarios in which the implicit substitution costs differ~across clusters. Hence, the \textsf{UU} model is analogous to the hierarchical Ward\textsubscript{$p$} algorithm \citep{deAmorim2015}, in the sense of having cluster-specific feature weights (albeit with no tuning required).

Given the role played by $\lambda$ when it takes the value $0$, whereby the state distribution is uniform, it is convenient and natural to include a noise component (denoted by \textsf{N}),~whose single precision parameter is fixed to $0$, to robustify inference by capturing deviant cases~and minimising their deleterious effects on parameter estimation for the other, more defined~\mbox{clusters}. Adding this extension to each of the $4$ models above, regardless of how precision parameters are otherwise specified, completes the MEDseq model family with the \textsf{CCN}, \textsf{UCN}, \textsf{CUN}, and \textsf{UUN} models. When $G=1$, the \textsf{CC}, \textsf{CU}, and \textsf{CCN} models can be fitted. When $G=2$, the \textsf{UCN} and \textsf{UUN} models are equivalent to the \textsf{CCN} and \textsf{CUN} models, respectively, as there is only one non-noise component. As the noise component arises naturally from restricting the parameter space, we consider the noise component as one of the $G$ components, denoted hereafter with the subscript $0$. 
All $8$ model types are summarised further in Appendix \ref{Section:ParameterCount}.\enlargethispage{\baselineskip}

\subsection[Incorporating Covariates]{Incorporating Covariates}
\label{Section:Gating}

We now illustrate how to incorporate the available covariate information into the clustering process, both to guide the construction of the clusters and to find the typical trajectories which can be best predicted by covariates. As is typical for model-based clustering analyses, the data are augmented in MEDseq models by introducing a latent cluster membership indicator vector $\smash{\mathbf{z}_i=\left(z_{i,1},\ldots,z_{i,G}\right)^\top}$, where $\smash{z_{i,g}=1}$ if observation $i$ belongs to cluster $g$ and $\smash{z_{i,g}=0}$ otherwise. The MEDseq approach can be easily extended to incorporate the possible effects of covariates on the assignments of sequences to clusters by allowing the covariates to influence the distribution of the latent variable~$\smash{\mathbf{z}_i}$. Thus, such covariates are interpreted differently from those used to define the sampling weights, if any.

The inclusion of covariates is achieved under the mixture of experts framework \citep{Jacobs1991, GormleySchnatter2019}, by extending the mixture model to allow the mixing proportions for observation $i$ to depend on covariates $\smash{\mathbf{x}_i}$. This, rather than having covariates enter through the component distributions, is particularly attractive, as the interpretation of the remaining component-specific parameters is the same as it would be under a model without covariates. For example, in the case of the \textsf{CC} MEDseq model\vspace{-0.75ex}
\[f\negthinspace\left(\mathbf{s}_i\given \boldsymbol{\theta}_1,\ldots,\boldsymbol{\theta}_G,\lambda,\boldsymbol{\beta}_1,\ldots,\boldsymbol{\beta}_G, \mathbf{x}_i,\mathrm{d_H}\right)^{w_i} = \left\lbrack\sum_{g=1}^G\tau_g\negthinspace\left(\mathbf{x}_i\given\boldsymbol{\beta}_g\right)\frac{\exp\negthinspace\left(-\lambda\mathrm{d_H}\negthinspace\left(\mathbf{s}_i,\boldsymbol{\theta}_g\right)\right)}{\left(\left(v-1\right)e^{-\lambda} + 1\right)^T}\right\rbrack^{w_i}\negmedspace,\vspace{-0.75ex}\]
where the mixing proportions $\smash{\tau_g\negthinspace\left(\mathbf{x}_i\given \boldsymbol{\beta}_g\right)}$ (henceforth $\smash{\tau_g\negthinspace\left(\mathbf{x}_i\right)}$, for simplicity) are referred to as the `gating network', with $\smash{\tau_g\negthinspace\left(\mathbf{x}_i\right) > 0}$ and $\smash{\sum_{g=1}^G\tau_g\negthinspace\left(\mathbf{x}_i\right)=1}$, as usual, and $\smash{\boldsymbol{\beta}_1,\ldots,\boldsymbol{\beta}_G}$ are the gating network regression parameters. Such a model can be seen as a conditional mixture model \citep{Bishop2006} because, given the covariates $\smash{\mathbf{x}_i}$, the distribution of the sequences is a finite mixture model under which $\smash{\mathbf{z}_i}$ has a multinomial distribution with a single trial and probabilities equal to $\smash{\tau_g\negthinspace\left(\mathbf{x}_i\right)}$. The distance-based $k$-medoids algorithm, though closely related (see Section \ref{Section:INIT}), does not accommodate the inclusion of gating covariates in this~way.

Incorporating covariates in `hard' clustering algorithms using MLR, as per \citet{mvad2002}, has been criticised because the hard assignment of extraneous cases can negatively impact internal cluster cohesion and the MLR coefficient estimates \citep{Piccarreta2019}. An advantage of the noise component in MEDseq models is that it captures uniformly distributed sequences that deviate from those in the other, more defined clusters. Filtering outliers in this way lessens their impact on the non-noise gating network coefficients, thereby enabling more accurate inference and improving the interpretability of the effects of the covariates. Moreover, the `soft' partition obtained under the model-based paradigm allows the cluster membership probabilities for sequences lying on the boundary between two neighbouring clusters to be quantified and the effect of such sequences on the gating network coefficients to be mitigated.

As per \citet{Murphy2020}, the \textsf{CCN}, \textsf{UCN}, \textsf{CUN}, and \textsf{UUN} models which include an explicit noise component can be restricted to having covariates only influence the mixing proportions for the non-noise components, with all observations therefore assumed to have equal probability of belonging to the uniform noise component (i.e. by replacing $\smash{\tau_0\negthinspace\left(\mathbf{x}_i\right)}$ with $\smash{\tau_0}$). We refer to the former setting as the gated noise (GN) setting and to the latter as the non-gated noise (NGN) setting. The NGN setting is the more parsimonious one, makes more clear the distinction between EDM components and the uniform component, and is particularly apt when $\tau_0$ is expected to be small and/or the sequences are expected to overwhelm the gating covariate(s) in determining which cases are noise. Gating covariates can only be included when $G \geq 2$ under the GN setting or when there are $2$ or more non-noise components under the NGN setting.

\section[Model Estimation]{Model Estimation}
\label{Section:Estimation}

This section describes our model-fitting approach and some implementation issues that arise in practice. Specifically, Section \ref{Section:ECM} outlines the ECM algorithm employed for parameter estimation, Section \ref{Section:INIT} discusses the initialisation thereof with reference to the similarities between MEDseq models and the $k$-medoids and $k$-modes \citep{Huang1997} algorithms, and the issues of model selection, covariate selection, and model validation are treated in Section~\ref{Section:Selection}.

\subsection[Model Fitting via ECM]{Model Fitting via ECM}
\label{Section:ECM}

Parameter estimation is greatly simplified by the existence of a closed-form expression for the normalising constant for MEDseq models based on the Hamming or weighted Hamming distances. We focus on maximum (pseudo) likelihood estimation using a simple variant of the EM algorithm \citep{Dempster1977}. For simplicity, model fitting details are described chiefly for the \textsf{CC} MEDseq model with sampling weights and gating covariates. Additional details for other model types are deferred to Appendix \ref{Section:AllSteps}; so, too, are technical details pertaining to estimation of the precision parameter(s). The complete data pseudo likelihood for the \textsf{CC} model is given by\vspace{-1ex}\enlargethispage{\baselineskip}
\begin{equation*}
\mathcal{L}_c^{\mathbf{w}}\negthinspace\left(\boldsymbol{\theta}_1,\ldots,\boldsymbol{\theta}_G,\lambda,\boldsymbol{\beta}_1,\ldots,\boldsymbol{\beta}_G\given\mathbf{S},\mathbf{X},\mathbf{Z},\mathbf{w},\mathrm{d_H}\right) = \prod_{i=1}^{n}\left\lbrack\prod_{g=1}^{G}\left(\tau_g\big(\mathbf{x}_i\big)\frac{\exp\negthinspace\left(-\lambda\mathrm{d_H}\negthinspace\left(\mathbf{s}_i,\boldsymbol{\theta}_g\right)\right)}{\left(\left(v-1\right)e^{-\lambda} + 1\right)^T}\right)^{z_{i,g}}\right\rbrack^{w_i}\negmedspace,
\end{equation*}
and the complete data pseudo log-likelihood hence has the form:\vspace{-1ex}
\begin{equation}
\begin{alignedat}{2}
\begin{split}
\hspace{-1.5mm}\ell_c^{\mathbf{w}}\negthinspace\left(\boldsymbol{\theta}_1,\ldots,\boldsymbol{\theta}_G,\lambda,\boldsymbol{\beta}_1,\ldots,\boldsymbol{\beta}_G\given\mathbf{S},\mathbf{X},\mathbf{Z},\mathbf{w},\mathrm{d_H}\right)= \sum_{i=1}^{n}\sum_{g=1}^{G}z_{i,g}w_i\left\lbrack\right.&\negthinspace\log\tau_g\big(\mathbf{x}_i\big) - \lambda\mathrm{d_H}\negthinspace\left(\mathbf{s}_i,\boldsymbol{\theta}_g\right)-\label{eq:cdllCC}\\[-1.125em] &\left.\negmedspace T\log\negthinspace\left(\left(v-1\right)e^{-\lambda}+1\right)\right\rbrack\negthinspace.
\end{split}
\end{alignedat}
\vspace{-1ex}
\end{equation}
Under this model, the distribution of $\smash{\mathbf{s}_i}$ depends on the latent cluster membership variable $\smash{\mathbf{z}_i}$, which in turn depends on covariates $\smash{\mathbf{x}_i}$, while $\smash{\mathbf{s}_i}$ is independent of $\smash{\mathbf{x}_i}$ conditional on $\smash{\mathbf{z}_i}$.

The iterative algorithm for MEDseq models follows in a similar manner to that for standard mixture models. It consists of an E-step (expectation) which replaces for each observation the missing data $\smash{\mathbf{z}_i}$ with their expected values $\smash{\smash{\widehat{\mathbf{z}}_i}}$, which sum to $1$, followed by a M-step (maximisation), which maximises the expected complete data pseudo log-likelihood. The M-step consists of a series of conditional maximisation (CM) steps in which each parameter is maximised individually, conditional on the other parameters remaining fixed. Hence, model fitting is in fact conducted using an expectation conditional maximisation (ECM) algorithm \citep{Meng1993}. Aitken's acceleration criterion is used to assess convergence of the non-decreasing sequence of weighted pseudo log-likelihood estimates \citep{Boehning1994,McNicholas2010c}. Parameter estimates produced on convergence achieve at least a local maximum of the pseudo likelihood function. Upon convergence, cluster memberships are estimated via the maximum \emph{a posteriori} (MAP) classification, i.e. cases are assigned to the cluster $g$ to which they most probably belong via $\smash{\textrm{MAP}\negthinspace\left(\mathbf{\widehat{z}}_i\right)=\argmax_{g\,\in\,\left\{1,\ldots,G\right\}}\big(\widehat{z}_{i,g}\big)}$.

The E-step (with similar expressions when $\lambda$ is unconstrained across clusters and/or time points) involves computing expression \eqref{eq:Estep}, where $(m+1)$ is the current iteration number:\vspace{-0.5ex}
\begin{equation}
\widehat{z}_{i,g}^{\left(m+1\right)} = \mathbb{E}\negthinspace\left(\negthinspace z_{i,g}\negthinspace\bigm|\negthinspace \mathbf{s}_i,\mathbf{x}_i,\widehat{\boldsymbol{\theta}}_g^{\left(m\right)}\negthinspace, \widehat{\lambda}^{\left(m\right)}\negthinspace, \widehat{\boldsymbol{\beta}}_g^{\left(m\right)}\negthinspace,w_i,\mathrm{d_H}\negthinspace\right)= \frac{\widehat{\tau}_g^{\left(m\right)}\negthinspace\big(\mathbf{x}_i\big)f\big(\mathbf{s}_i\negthinspace\bigm|\negthinspace \widehat{\boldsymbol{\theta}}_g^{\left(m\right)}\negthinspace,\widehat{\lambda}^{\left(m\right)}\negthinspace,\mathrm{d_H}\big)}{\sum_{h=1}^{G}\negthinspace\widehat{\tau}_{h}^{\left(m\right)}\negthinspace\big(\mathbf{x}_i\big) f\big(\mathbf{s}_i\negthinspace\bigm|\negthinspace \widehat{\boldsymbol{\theta}}_{h}^{\left(m\right)}\negthinspace,\widehat{\lambda}^{\left(m\right)}\negthinspace,\mathrm{d_H}\big)}.\label{eq:Estep}
\vspace{-0.5ex}
\end{equation}
Note that the weights $\smash{w_i}$ appear in neither the numerator nor the denominator, leaving the E-step unchanged regardless of the inclusion or exclusion of weights.

Subsequent subsections describe the CM-steps for estimating the remaining parameters in the model. These individual CM-steps rely on the current estimates $\widehat{\mathbf{Z}}^{\left(m+1\right)} = \big(\widehat{\mathbf{z}}_1^{\left(m+1\right)},\ldots,\widehat{\mathbf{z}}_n^{\left(m+1\right)}\big)$ to provide estimates of the gating network regression coefficients $\smash{\widehat{\boldsymbol{\beta}}_g^{\left(m+1\right)}}$, and hence the mixing proportion parameters $\smash{\widehat{\tau}^{\left(m+1\right)}_g\negthinspace\left(\mathbf{x}_i\right)}$, as well as the central sequence(s) $\smash{\widehat{\boldsymbol{\theta}}{\mathstrut}^{\left(m+1\right)}_g}$ and component precision parameter(s) $\smash{\widehat{\lambda}^{\left(m+1\right)}}$, though technical details for the latter, as they are the element which distinguishes the various MEDseq model types, are deferred to Appendix \ref{Section:AllSteps}. It is clear from \eqref{eq:cdllCC} that the sampling weights can be accounted for by simply multiplying every $\smash{\widehat{\mathbf{z}}_i^{\left(m+1\right)}}$ by the corresponding weight $\smash{w_i}$. Conversely, in the CM-steps which follow, corresponding formulas for unweighted MEDseq models can be recovered by~\mbox{replacing} $\smash{\widehat{z}_{i,g}^{\left(m+1\right)}w_i}$  with $\smash{\widehat{z}_{i,g}^{\left(m+1\right)}}$. 

\subsubsection[Estimating the Gating Network Coefficients]{Estimating the Gating Network Coefficients}
\label{Section:gatingcoeff}

The portion of \eqref{eq:cdllCC} corresponding to the gating network, given by $\smash{\sum_{i=1}^{n}\sum_{g=1}^{G}z_{i,g}w_i\log\tau_g\negthinspace\left(\mathbf{x}_i\right)}$,
is of the same form as a MLR model with weights given by $\smash{w_i}$\,, here written with component $1$ as the baseline reference level for identifiability reasons:
\[\log\frac{\tau_g\negthinspace\left(\mathbf{x}_i\right)}{\tau_1\negthinspace\left(\mathbf{x}_i\right)}=\log\frac{\Pr\negthinspace\left(z_{i,g}=1\right)}{\Pr\negthinspace\left(z_{i,1}=1\right)}=\widetilde{\mathbf{x}}_i\boldsymbol{\beta}_g\:\forall\:g\geq 2,~\mbox{with}~\boldsymbol{\beta}_1=\left(0,\ldots,0\right)^\top\negmedspace,\]
where $\smash{\widetilde{\mathbf{x}}_i = \left(1, \mathbf{x}_i\right)}$. Thus, methods for fitting such models, with $\smash{\widehat{\mathbf{Z}}^{\left(m+1\right)}}$ as the response, can be used to estimate the gating network regression parameters $\smash{\widehat{\boldsymbol{\beta}}_g^{\left(m+1\right)}}$. As closed-form updates are unavailable for MLR coefficients, due to the nonlinear numerical optimisation involved, this step merely increases (rather than maximises) the expectation of this term. However, the monotonicity of the sequence of pseudo log-likelihood estimates is preserved and convergence is still guaranteed.  Subsequently, the mixing proportions are given by\vspace{-0.67ex}
\[\widehat{\tau}_g^{\left(m+1\right)}\negthinspace\left(\mathbf{x}_i\right) = \frac{\exp\negthinspace\big(\widetilde{\mathbf{x}}_i\widehat{\boldsymbol{\beta}}_g^{\left(m+1\right)}\big)}{\sum_{h=1}^G\exp\negthinspace\big(\widetilde{\mathbf{x}}_i\widehat{\boldsymbol{\beta}}_h^{\left(m+1\right)}\big)}.\vspace{-0.67ex}\]
Conversely, $\boldsymbol{\tau}$ is estimated exactly via $\widehat{\tau}_g^{\left(m+1\right)} = n^{-1} \sum_{i=1}^n\widehat{z}_{i,g}^{\left(m+1\right)}w_i$ when there are no gating covariates. Since $\sum_{i=1}^nw_i=n$, this is simply the weighted mean of the $g$-th column of the matrix $\smash{\widehat{\mathbf{Z}}^{\left(m+1\right)}}$. However, $\boldsymbol{\tau}$ can also be constrained to be equal (i.e. $\tau_g=\nicefrac{1}{G}\:\forall\:g$) across clusters. Thus, situations where $\tau_{i,g}=\tau_g\negthinspace\left(\mathbf{x}_i\right)$, $\tau_{i,g}=\tau_{g}$, or $\tau_{i,g}=\nicefrac{1}{G}$ are accommodated.

The standard errors of the gating network's MLR at convergence are not a valid means of assessing the uncertainty of the coefficient estimates as the cluster membership probabilities are estimated rather than fixed and known. Therefore, we adapt the weighted likelihood bootstrap (WLBS) of \citet{OHagan2019} to the MEDseq setting. This is implemented by multiplying the sampling weights $\mathbf{w}$ by draws from an $n$-dimensional symmetric uniform Dirichlet distribution and refitting the MEDseq model. To ensure stable estimation of the standard errors, $B=1000$ such samples are used here. To ensure rapid convergence and to circumvent label-switching problems, the estimated $\widehat{\mathbf{Z}}$ matrix from the original model is used to initialise the ECM algorithm for each sample with new likelihood weights. Finally, the standard errors of the gating network coefficients across the $B$ samples are obtained. Although this approach does not produce fully valid variance estimates when there are sampling weights which arise from stratified designs, we adopt the WLBS in what follows in order to provide approximate standard errors. This issue is particularly pronounced when the probability of being included in the sample depends on quantities being modelled. This concern provides additional justification for the aforementioned removal of the Grammar and Location covariates from our analysis.

\subsubsection[Estimating the Central Sequences]{Estimating the Central Sequences}
\label{Section:theta}

The location parameter $\boldsymbol{\theta}$ is sometimes referred to as the Fr{\'{e}}chet mean or the central sequence. The $k$-medoids/PAM algorithm, which is closely related to the MEDseq models with certain restrictions imposed (see Section \ref{Section:INIT}), fixes the estimate of $\smash{\widehat{\boldsymbol{\theta}}_g}$ to be the medoid of cluster $g$ \citep{Kaufman1990}, i.e. the observed sequence $\smash{\mathbf{s}_i\in\mathbf{S}}$ with minimum weighted distance from the others currently assigned to the same cluster. This estimation approach is especially quick as the Hamming distance matrix for the observed sequences is pre-computed. Notably, this greedy search strategy may fail to find the optimum solution.

However, for a $G=1$ unweighted EDM based on the Hamming distance, the maximum likelihood estimate (MLE) of $\boldsymbol{\theta}$ is given simply by the modal sequence, meaning that each $\smash{\widehat{\theta}_t}$ is independently given by the most frequent state at the $t$-th time point. This is intuitive when $\smash{\mathrm{d_H}\negthinspace\left(\mathbf{s}_i, \boldsymbol{\theta}\right)}$ is expressed as $\smash{T - \sum_{t=1}^T \indicator{s_{i,t} = \theta_t}}$, as $\smash{\widehat{\boldsymbol{\theta}}}$ maximises the number of element-wise agreements. Thus, the parameter has a natural interpretation. For more complicated distance metrics, the first-improvement algorithm \citep{Hoos2004} or a genetic algorithm could be used to estimate $\boldsymbol{\theta}$. 
Notably, the modal sequence need not be an observed sequence in $\mathbf{S}$. It is also notable that any $\smash{\widehat{\theta}_t}$ may be non-unique under any of the proposed estimation strategies. Such ties, if any, are broken at random.\enlargethispage{\baselineskip}

For $G>1$, under the ECM framework, central sequence position estimates $\smash{\widehat{\theta}_{g,t}^{\left(m+1\right)}}$ are given by $\smash{\argmax_{\vartheta\,\in\,\bm{\mathcal{V}}_t}\negthinspace\big(\sum_{i=1}^n\widehat{z}_{i,g}^{\left(m+1\right)}w_i\indicator{s_{i,t} = \vartheta}\big)}$, where $\smash{\bm{\mathcal{V}}_t}$ is the subset of $\smash{v_t\leq v}$ states observed at time point $t$ across all cases. As this expression is independent of the precision parameter(s), it holds for all MEDseq model types, including those based on weighted Hamming distance variants. Thus, $\smash{\widehat{\boldsymbol{\theta}}_{g}^{\left(m+1\right)}}$ is similarly estimated easily and exactly via a weighted mode (much like $k$-modes), whereby each $\smash{\widehat{\theta}_{g, t}^{\left(m+1\right)}}$ is given by the category corresponding to the maximum of the sum of the weights $\smash{\widehat{z}_{i,g}^{\left(m+1\right)}w_i}$ associated with each of the $\smash{v_t}$ observed state values. Similarly, the central sequence under a weighted $G=1$ model is also estimated via a weighted mode, with the weights given only by $\smash{w_i}$. Notably, to estimate the central sequences for a MEDseq model of any type without sampling weights, one need only remove $\smash{w_i}$ from these terms. Note also that $\smash{\boldsymbol{\theta}_0}$ does not need to be estimated for models with an explicit noise component as it does not contribute to the likelihood.

\subsection[ECM Initialisation and Comparison to $k$-medoids]{ECM Initialisation and Comparison to $k$-medoids}
\label{Section:INIT}

MEDseq models share relevant features with the PAM algorithm. Both consider sequences from a holistic perspective and both rely on distances to a cluster centroid. However,~PAM treats the matrix of pairwise distances between sequences as a pre-computed input, while under MEDseq models the distances to the centroids (and the costs which define the~\mbox{distance} metric) are recomputed at each iteration, with the sequences themselves as input. Otherwise, compared to PAM based on the Hamming distance, MEDseq models differ only~in~that i) $\smash{\boldsymbol{\theta}_g}$ is estimated by the modal sequence rather than the medoid, ii) $\boldsymbol{\tau}$ is estimated, or dependent on covariates via $\smash{\tau_g\negthinspace\left(\mathbf{x}_i\right)}$, rather than constrained to be equal, iii) $\lambda$ is free to~vary~across clusters and/or time points, rather than being implicitly set to $1$, iv) a noise component~can be included, and v) the ECM algorithm rather than the classification EM algorithm (CEM; \citealp{Celeux1992}) is used. The CEM algorithm employed by PAM uses \emph{hard} assignments $\smash{\widetilde{z}_{i,g}}$, computed in its C-step, such that $\smash{\widetilde{z}_{i,g}^{\left(m+1\right)} = 1~\mbox{if}~g= \textrm{MAP}\big(\widehat{\mathbf{z}}_i^{\left(m+1\right)}\big)}$ and $\smash{\widetilde{z}_{i,g}^{\left(m+1\right)} = 0}$ otherwise, for which the denominator in \eqref{eq:Estep} need not be evaluated.~\mbox{Overall}, relaxing the constraints on $\boldsymbol{\tau}$ and $\lambda$ combines with the use of \emph{soft} assignments to help~prevent ties, for both $\smash{\widehat{\theta}_t}$ and the cluster allocations, which often plague clustering methods for categorical~data.

Thus, a \textsf{CC} model fitted by CEM, with $\lambda=1$, equal mixing proportions, and the central sequences estimated by the medoid rather than the modal sequence, is equivalent to $k$-medoids based on the Hamming distance. We leverage these similarities by applying $k$-medoids to the Hamming distance matrix in order to initialise the ECM algorithm with `hard' starting values for the allocation matrix $\mathbf{Z}$. In particular, we rely on a weighted version of PAM available in the \textsf{R} package \texttt{WeightedCluster} \citep{WeightedCluster2013}, itself initialised using Ward's hierarchical clustering. The more closely related $k$-modes algorithm \citep{Huang1997} is not used, as case-weighted implementations are currently unavailable. In any case, our strategy is less computationally onerous than using multiple random starts. Moreover, our experience suggests that the ECM algorithm converges quickly when our initialisation strategy is adopted and that a great many number of random starts are required in order to achieve comparable performance. For models with an explicit noise component, an initial guess of the prior probability $\tau_0$ that observations are noise is required. Allocations are then initialised, assuming the last component is the one associated with $\smash{\lambda_g=0}$, by multiplying the initial ($G-1$)-column $\mathbf{Z}$ matrix by $\smash{1-\tau_0}$ and appending a column in which each entry is $\smash{\tau_0}$. We caution that the initial $\smash{\tau_0}$ should not be too large.\enlargethispage{\baselineskip}

\subsection[Model Selection and Validation]{Model Selection and Validation}
\label{Section:Selection}

In contrast to heuristic clustering approaches like $k$-medoids and Ward's hierarchical method, the model-based paradigm facilitates principled model-selection using likelihood-based information criteria. In the MEDseq setting, the notion of model selection refers to identifying the optimal number of components $G$ in the mixture and finding the best MEDseq model type in terms of constraints on the precision parameters. Variable selection on the subset of covariates included in the gating network can also improve the fit. For a given set of covariates, one would typically evaluate all model types over a range of $G$ values and choose simultaneously both the model type and $G$ value according to some criterion. Thereafter, different fits with different covariates can be compared according to the same criterion.

The Bayesian Information Criterion (BIC; \citealt{Schwarz1978}) includes a penalty term which depends on the number of free parameters $k$ in the model. The parameter counts can be deceptive for MEDseq models. In particular, regarding the estimation of $\smash{\widehat{\theta}_{g, t}}$, we note that identifying the modal state for a given time point implicitly involves estimating occurrence probabilities for $\smash{\left(v_t-1\right)}$ states and then selecting the most common. This is accounted for in Appendix \ref{Section:ParameterCount}, wherein the number of free parameters in under each MEDseq model type is summarised. We also note that the penalty $k\log n$ is applied to the maximum \emph{pseudo} log-likelihood estimate in the sample-weighted setting \citep{Xu2013}.

Beyond its use in identifying the optimal $G$ and precision parameter settings, the~BIC is also employed in greedy stepwise selection algorithms in order to guide the inclusion/exclusion of relevant gating covariates. We propose a bi-directional search strategy in which each step can potentially consist of adding or removing a non-noise component or adding or removing a covariate. Interaction terms are not considered. Every potential action is evaluated over all possible model types at each step, rather than considering changing the model type as an action in itself. Changing the gating covariates or changing the number of components can affect the model type, as observed by \citet{Murphy2020}. While this makes the stepwise search more computationally intensive, it is less likely to miss optimal models as it explores the model space. For steps involving both gating covariates and a noise component, models with both the GN and NGN settings can be evaluated and potentially selected.

A backward stepwise search starts from the model, with all covariates included, considered optimal in terms of the number of components $G$ and the MEDseq model type. On the other hand, a forward stepwise search uses the optimal model with no covariates included as its starting point. In both cases, the algorithm accepts the action yielding the highest increase in the BIC at each step. The computational benefits of upweighting unique cases and discarding redundant cases are stronger for the forward search, as early steps with fewer covariates are likely to have fewer unique cases across sequence patterns and covariates.

As a means of validating the model chosen by BIC, we turn to silhouette analysis to assess the quality of the clustering in terms of internal cohesion, where high cohesion indicates high between-cluster distances and strong within-cluster homogeneity. Typically, the silhouette width is defined for clustering methods which produce a `hard' partition \citep{Rousseeuw1987}, and the average silhouette width (ASW) or weighted average silhouette width (wASW; \citealt{WeightedCluster2013}) is used as a model selection criterion. However, \citet{Menardi2011} introduces the density-based silhouette (DBS) for model-based clustering methods. This allows the `soft' assignment information to be used, which is discarded when using the MAP assignments in the computation of the wASW. The empirical DBS for observation $i$ is given by\vspace{-1.25ex}
\begin{equation}
\widehat{dbs}_i = \frac{\log\negthinspace\left(\dfrac{\widehat{z}_i^0}{\widehat{z}_i^1}\right)}{\max_{h\,\in\,\left\{1,\ldots,n\right\}}\left(\left\lvert\log\negthinspace\left(\dfrac{\widehat{z}_h^0}{\widehat{z}_h^1}\right)\right\rvert\right)}.\label{eq:dbs}\vspace{-1.25ex}
\end{equation}
\indent As observations are assigned to clusters via the MAP classification, $\smash{\widehat{dbs}_i}$ is proportional to the log-ratio of the posterior probability associated with the MAP assignment of observation $i$ (denoted by $\smash{\widehat{z}_i^0}$) to the maximum posterior probability that the observation belongs to another cluster (denoted by  $\smash{\widehat{z}_i^1}$). Use of the MAP classification implies $\smash{0 \leq \widehat{dbs}_i \leq 1\:\forall\:i}$, with high values indicating a well-clustered data point. Ultimately, the mean or the median $\smash{\widehat{dbs}}$ value can be used as a global quality measure, albeit with two modifications. Firstly, we identify a set of crisply assigned observations having $\smash{\widehat{z}_i^1}$ lower than a tolerance parameter $\epsilon$, here set equal to $\smash{10^{-100}}$. These observations are given $\smash{\widehat{dbs}_i}$ values of $1$ and are excluded from the computation of the maximum in the denominator of \eqref{eq:dbs} for reasons of numerical stability. Secondly, we account for the sampling weights by computing a \emph{weighted} mean~\mbox{density-based} silhouette criterion (wDBS). While neither the wDBS nor wASW are defined for $G=1$, unlike the BIC, they are not employed here as model selection criteria. These silhouette summary measures are used only to validate MEDseq clustering solutions and to facilitate comparisons with other methods in Section \ref{Section:Comparison}. Higher values are preferred for both criteria.

\section[Analysing the MVAD Data]{Analysing the MVAD Data}
\label{Section:Results}

Results of fitting MEDseq models to the weighted MVAD data are provided in Section \ref{Section:MVADresults}. All results were obtained via our purpose-built \textsf{R} package \texttt{MEDseq} \citep{MEDseqR2021}. The impact of discarding the sampling weights is also studied. A comparison against other approaches, including hierarchical, partitional, and model-based clustering methods, is included in Section \ref{Section:Comparison}. A discussion of the insights gleaned from the solution obtained by the optimal MEDseq model is deferred to Section \ref{Section:Discussion}.\enlargethispage{\baselineskip}

\subsection[Application of MEDseq]{Application of MEDseq}
\label{Section:MVADresults}

\begin{figure}[H]
	\centering
	\includegraphics[width=0.7125\textwidth,keepaspectratio]{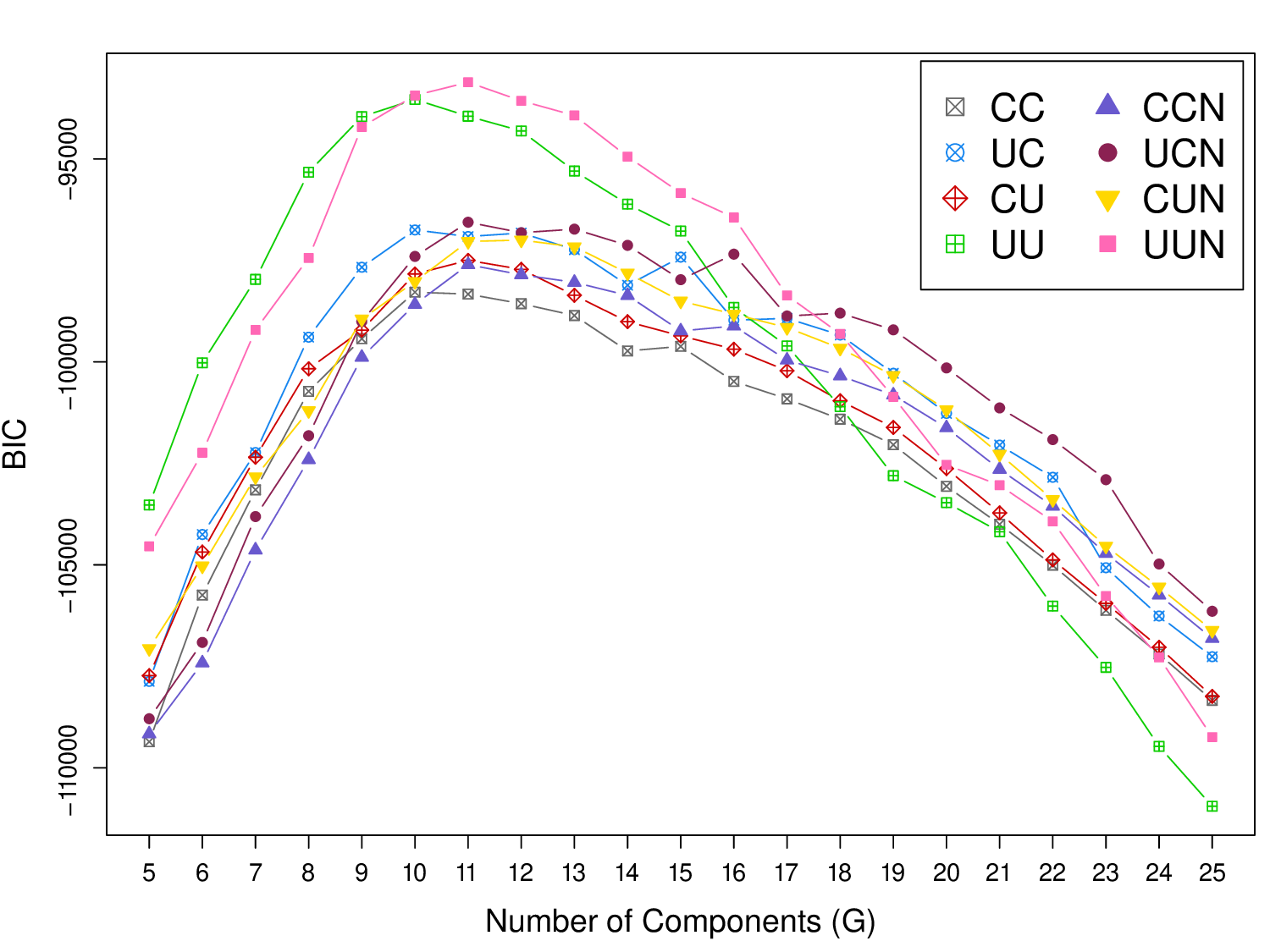}
	\caption{BIC values for all MEDseq model types, with weights and all covariates, for a range of $G$ values.}
	\label{Plot:MVADbic}
\end{figure}

Weighted MEDseq models are fit for $G=1,\ldots,25$, across all $8$ model types (where allowable), firstly with all covariates included in the gating network (again, where allowable). The noise components, where applicable, are treated using the NGN setting. Figure \ref{Plot:MVADbic} shows the behaviour of the BIC for these models. To better highlight the differences in BIC, lower values for $G < 5$ are not shown. Under these conditions, a $G=11$ \textsf{UUN} model is identified as optimal. The same model type and number of components are identified as optimal when the noise components are treated with the GN setting, and when the same analysis is repeated with no covariates at all.
\vspace{-1ex}
In refining the model further via greedy stepwise selection, both the forward search (see Table \ref{Table:mvadForwardStep}) and backward search (see Table \ref{Table:mvadBackwardStep}) thus begin with the same number of components and the same model type. As previously stated, covariates used to define the sampling weights are excluded in both cases. Notably, no step in either search elects to modify $G$ or the model type. Both searches converge to the same $G=11$ \textsf{UUN} model with only the single covariate `GCSE5eq' in the NGN gating network, though the search in the forward direction does so in fewer steps. Under this model, the probability of belonging to the noise component is constant and does not depend on the included covariate.\vspace{-1ex}
\begin{table}[H]
	\caption{Summary of the steps taken to improve the BIC in the forward direction.\label{Table:mvadForwardStep}}
	\centering
	\scriptsize
	\setlength{\tabcolsep}{4.25pt}
	\begin{tabular}[pos=center]{c c c c c c}
		\specialrule{.1em}{.01em}{.01em} 
		Optimal Step & $G$ & Model Type & Gating Covariates & Gating Type & BIC\\
		\hline\hline
		--- & $11$ & \textsf{UUN} &  & --- & $-93190.08$\\
		Add `GCSE5eq' & $11$ & \textsf{UUN} & GCSE5eq & NGN& $-92953.85$\\
		Stop & $11$ & \textsf{UUN} & GCSE5eq & NGN&$-92953.85$\\
		\specialrule{.1em}{.01em}{.01em} 
	\end{tabular}
\end{table}
\vspace{-1em}
\begin{table}[H]
	\caption{Summary of the steps taken to improve the BIC in the backward direction.\label{Table:mvadBackwardStep}}
	\centering
	\scriptsize
	\setlength{\tabcolsep}{4.25pt}
	\begin{tabular}[pos=center]{c c c c c c}
		\specialrule{.1em}{.01em}{.01em} 
		Optimal Step & $G$ & Model Type & Gating Covariates & Gating Type & BIC\\
		\hline\hline
		--- & $11$ & \textsf{UUN} & Catholic, FMPR, Funemp, GCSE5eq, Gender, Livboth   & NGN & $-93111.30$\\
		Remove `FMPR' & $11$ & 	\textsf{UUN} & Catholic, Funemp, GCSE5eq,  Gender, Livboth & NGN	& $-93068.09$\\
		Remove `Livboth' & $11$ &	\textsf{UUN} & Catholic, Funemp, GCSE5eq, Gender    & NGN& $-93025.73$\\
		Remove `Catholic' & $11$ & \textsf{UUN} & Funemp, GCSE5eq, Gender    & NGN 	& $-92994.32$\\
		Remove `Funemp' & $11$ & \textsf{UUN} & GCSE5eq, Gender    & NGN 	& $-92967.23$\\
		Remove `Gender' & $11$ & \textsf{UUN} & GCSE5eq& NGN 	& $-92953.85$\\
		Stop & $11$ & \textsf{UUN} & GCSE5eq & NGN	&	$-92953.85$\\
		\specialrule{.1em}{.01em}{.01em} 
	\end{tabular}
\end{table}
\vspace{-1ex}
Notably, there is little difference between the respective clusterings produced by the various models including no covariates, all covariates, and only GCSE5eq. Indeed, both the soft $\widehat{\mathbf{Z}}$ matrices and hard MAP assignments are almost identical between each pair of models; relative to the optimal model after stepwise selection, there are only $1$ and $2$ cases assigned to different clusters under equivalent models with no covariates and all covariates, respectively. Thus, the sequences themselves overwhelm the covariates and there is little confounding between the simultaneous roles of GCSE5eq under the optimal model in guiding both the construction of the clusters and their interpretation. Moreover, the parsimony afforded by discarding the other covariates simplifies the interpretation greatly. Thus, while adapting the `two-step' approach introduced for LCR \citep{Bakk2018} to the MEDseq setting may be of interest for other applications, the results for the MVAD data do not differ greatly from those presented in Section \ref{Section:Discussion}, as shown in Appendix \ref{Section:AllGating}.\enlargethispage{\baselineskip}

For completeness, the analysis above is repeated with the sampling weights discarded entirely and consideration given where appropriate to the two covariates used to  define $\mathbf{w}$. In doing so, identical inference is obtained on the model type; however, the results differ in terms of the optimal $G$ (now $10$), the uncovered partition, and the estimated model parameters. This is not surprising, as failure to account for $\mathbf{w}$ in the clustering produces biased estimates of the component-specific parameters and the cluster membership probabilities, as well as the gating network coefficients. Additionally, an extra gating covariate (Grammar) is included after stepwise selection in the unweighted analysis. However, the results are reasonably robust to a coarsening of the sequences; in repeating all analyses with the data subsetted into six-monthly intervals, similar inferences are again obtained. Notably, the ECM algorithm's runtime is not greatly reduced in doing so. Indeed, MEDseq models scale more poorly with $n$ (or, more specifically, the number of unique cases) rather than $T$ or $v$, as the number of (pseudo) likelihood evaluations required for large $n$ is more computationally expensive than the number of simple matching evaluations required for long sequences.

\subsection[Other Clustering Methods]{Other Clustering Methods}
\label{Section:Comparison}

To contrast the MEDseq results for the MVAD data with those obtained by other methods, we present a non-exhaustive comparison against some distance-based and some Markovian approaches. Regarding the former, we present only some common heuristic methods which treat the distance matrix as the input using distance metrics which are commonly adopted in the literature on life-course sequences, namely PAM and Ward's method based on the Hamming distance and OM. We note that fuzzy clustering offers an alternative distance-based perspective which also allows for soft assignments (see \citet{Durso2016} for an excellent overview), with further, separate extensions for incorporating covariates and including a noise cluster in \citet{Studer2018} and \citet{Durso2013}, respectively. However, this paradigm is not considered further, both for the sake of brevity and because case-weighted implementations are currently unavailable. LCA and LCR, fit via the \textsf{R} package \texttt{poLCA} \citep{Linzer2011}, are also excluded, as they encounter computational difficulties due to the explosion in the number of parameters for $\smash{G\geq3}$. Among the considered methods, only MEDseq and the distance-based methods can accommodate the sampling weights.

\begin{figure}[H]
	\centering
	\includegraphics[width=0.6\textwidth,keepaspectratio]{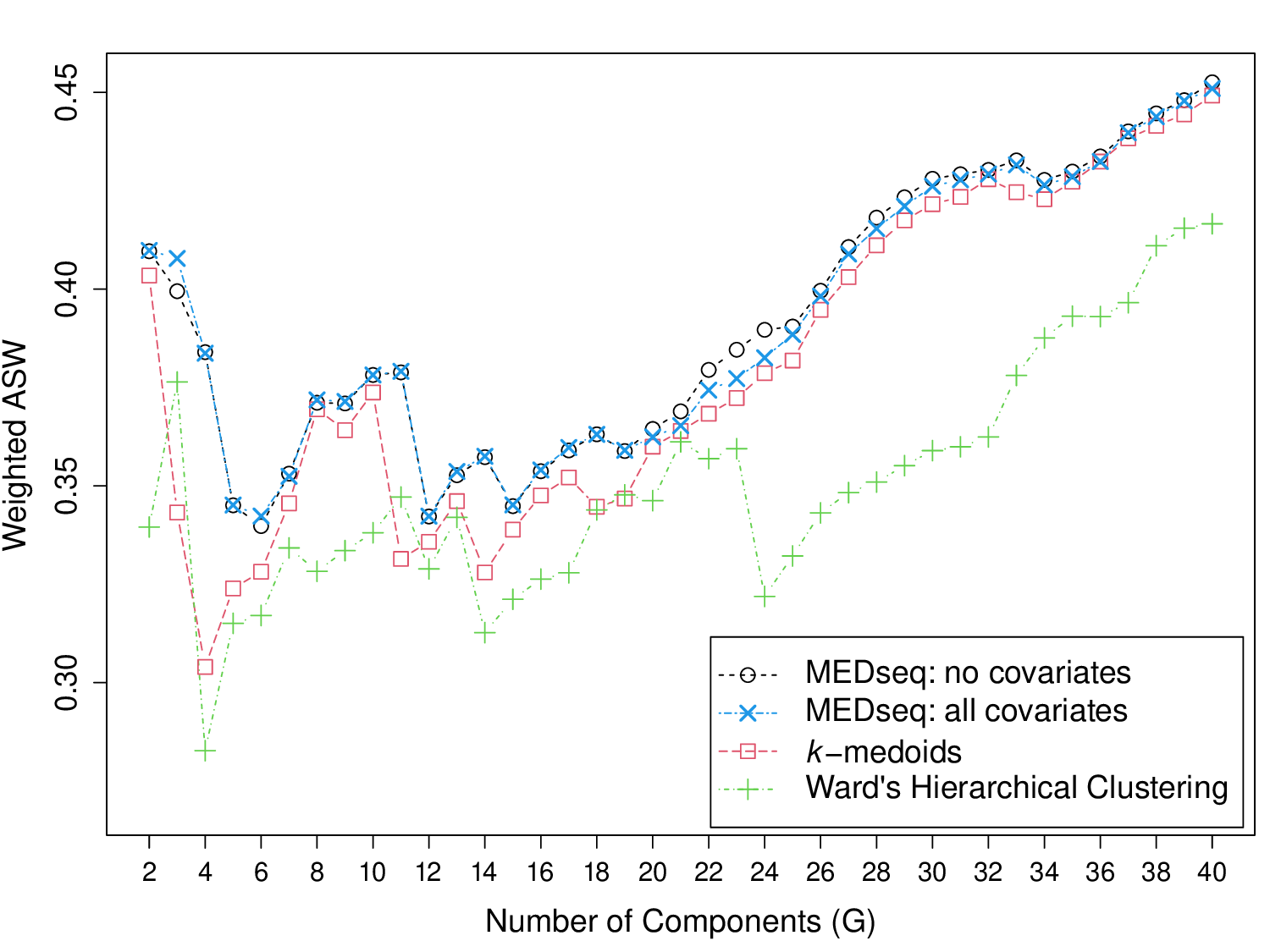}
	\caption{Values of the wASW measure, using Hamming distances, for the best MEDseq model type for each $G$ value with no covariates and all covariates. Corresponding values for weighted versions of $k$-medoids and Ward's hierarchical clustering based on the Hamming distance are also shown.\label{Plot:mvadWeightASW}}
\end{figure}

\begin{figure}[H]
	\centering
	\includegraphics[width=0.6\textwidth,keepaspectratio]{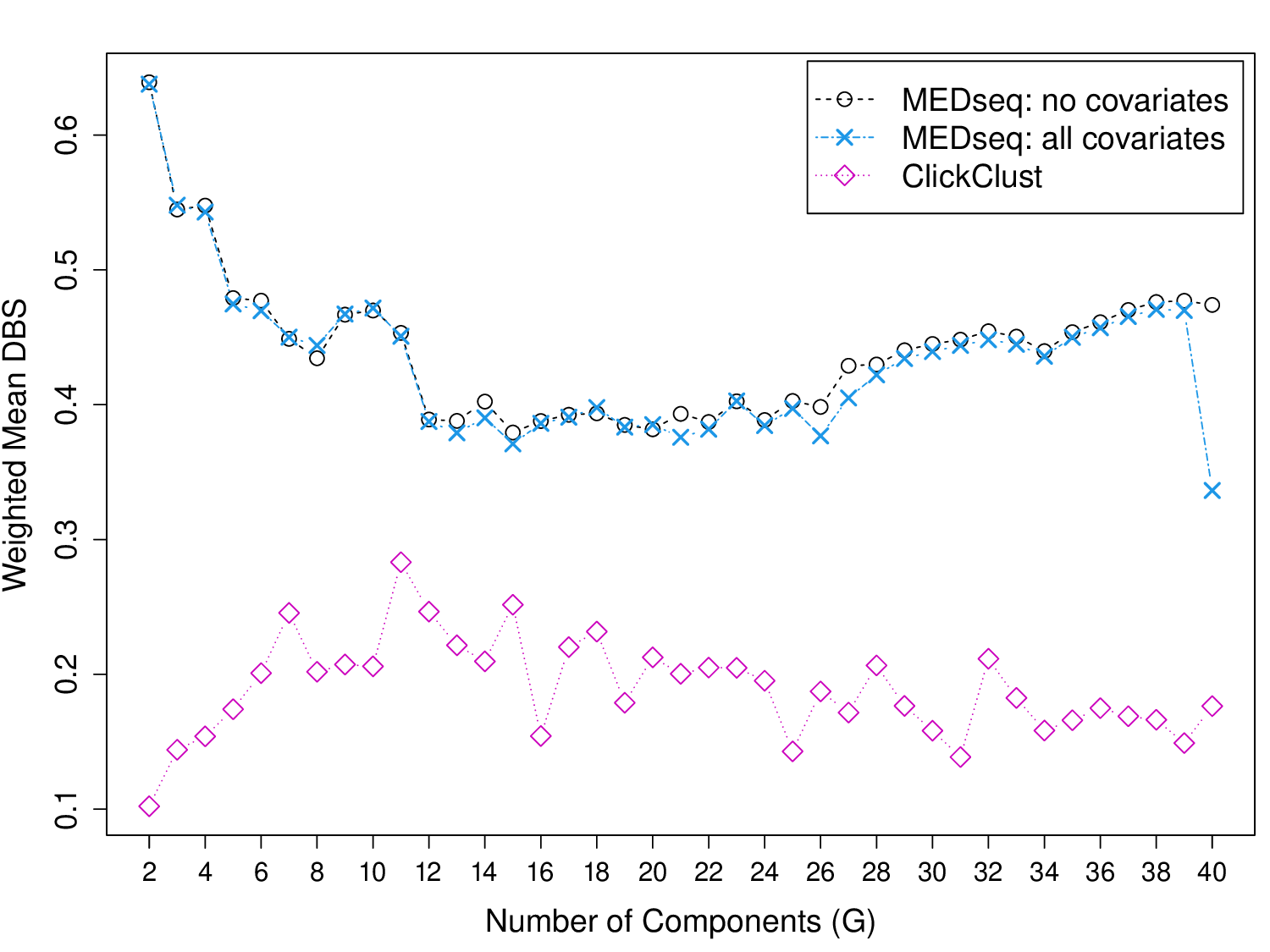}
	\caption{Values of the wDBS measure for the best MEDseq model type at each $G$ value with no covariates and all covariates. Corresponding values for the best \texttt{ClickClust} model are also shown.\label{Plot:mvadClickDBS}}
\end{figure}

Firstly, MEDseq models with no covariates and all covariates are compared against weighted versions of $k$-medoids, using the \textsf{R} package \texttt{WeightedCluster} \citep{WeightedCluster2013}, and Ward's hierarchical clustering. Here, $k$-medoids is itself initialised using Ward's method. Neither method can be compared to MEDseq models in terms of BIC or wDBS~\mbox{values},~as~they are not model-based and do not yield `soft' cluster membership probabilities, respectively. Thus, Figure \ref{Plot:mvadWeightASW} shows a comparison of wASW values using MAP classifications where necessary. Only the MEDseq model type (and gating network setting, for models with covariates) with the highest wASW for each $G$ value is shown, for clarity. Note that the wASW is computed using the observed Hamming distance matrix, which both comparators in Figure \ref{Plot:mvadWeightASW} utilise directly, while MEDseq models are only \emph{based on} the Hamming metric. Nonetheless, MEDseq models show superior or competitive performance across the majority of $G$ values. In particular, the optimal model identified after stepwise selection achieves wASW=$0.386$. The superior wASW values achieved by MEDseq models provide evidence that the proposed methodology, which embeds features of the distance-based approaches into a model-based setting, yields more compact and well-separated clusters. Notably, similar conclusions are drawn when OM --- with the same cost settings as used in \citet{mvad2002} --- is used in place of the Hamming distance for $k$-medoids and Ward's method.\vspace{-1em}\enlargethispage{\baselineskip}
\vspace{-1.5ex}
Secondly, finite mixtures with first-order Markov components, fit via the \textsf{R} package \texttt{ClickClust} \citep{ClickClust2016}, are also included in the comparison. This package allows the initial state probabilities to be either estimated or equal to $\smash{\nicefrac{1}{v}}$ for all categories; both scenarios are considered and other function arguments are set to their default values. The wASW values for the \texttt{ClickClust} models are not shown in Figure \ref{Plot:mvadWeightASW}; they are much lower than those of the other methods up to $G=5$ and turn negative thereafter. Though this implies inferior clustering behaviour for \texttt{ClickClust} models, the method also returns a $\smash{\widehat{\mathbf{Z}}}$ matrix of cluster membership probabilities. Hence, these models are also compared to MEDseq in terms of the wDBS measure in Figure \ref{Plot:mvadClickDBS}. Again, only the best model of each type is shown for each $G$ value; here, the MEDseq models again exhibit the best performance over the entire range. Notably, the optimal $G=11$ \textsf{UUN} MEDseq model with `GCSE5eq' in the gating network achieves wDBS=$0.455$. An advantage of \texttt{ClickClust} is that it allows sequences of unequal lengths, but this is not a concern for the MVAD data.\vspace{-1em}\enlargethispage{\baselineskip}
Thirdly, the \textsf{R} package \texttt{seqHMM} \citep{seqHMM2019} provides tools for fitting mixtures of hidden Markov models, with gating covariates influencing cluster membership probabilities. Such models allow cluster memberships to evolve over time, similar to mixed membership models \citep{Airoldi2014}. They thus cannot be directly compared to MEDseq models. However, we note that the \texttt{seqHMM} package provides a pre-fitted model for the MVAD data, with the first two months also discarded and no covariates. The model has $2$ clusters, with $3$ and $4$ hidden states, respectively, and achieves wDBS=$0.50$ and wASW=$0.23$. Otherwise identical \texttt{seqHMM} models, including either all covariates or only the GCSE5eq covariate chosen for the optimal MEDseq model via stepwise selection, both achieve wDBS=$0.47$ and wASW=$0.23$. Notably, these wDBS and wASW values are much worse than those for MEDseq models with $G=2$. Overall, the \texttt{ClickClust} and \texttt{seqHMM} results suggest that holistic approaches --- MEDseq models, in particular --- yield better clusterings than Markovian ones for the MVAD data.

\section[Discussion of the MVAD Results]{Discussion of the MVAD Results}
\label{Section:Discussion}

\begin{figure}[H]
	\centering
	\includegraphics[width=0.75\textwidth,keepaspectratio]{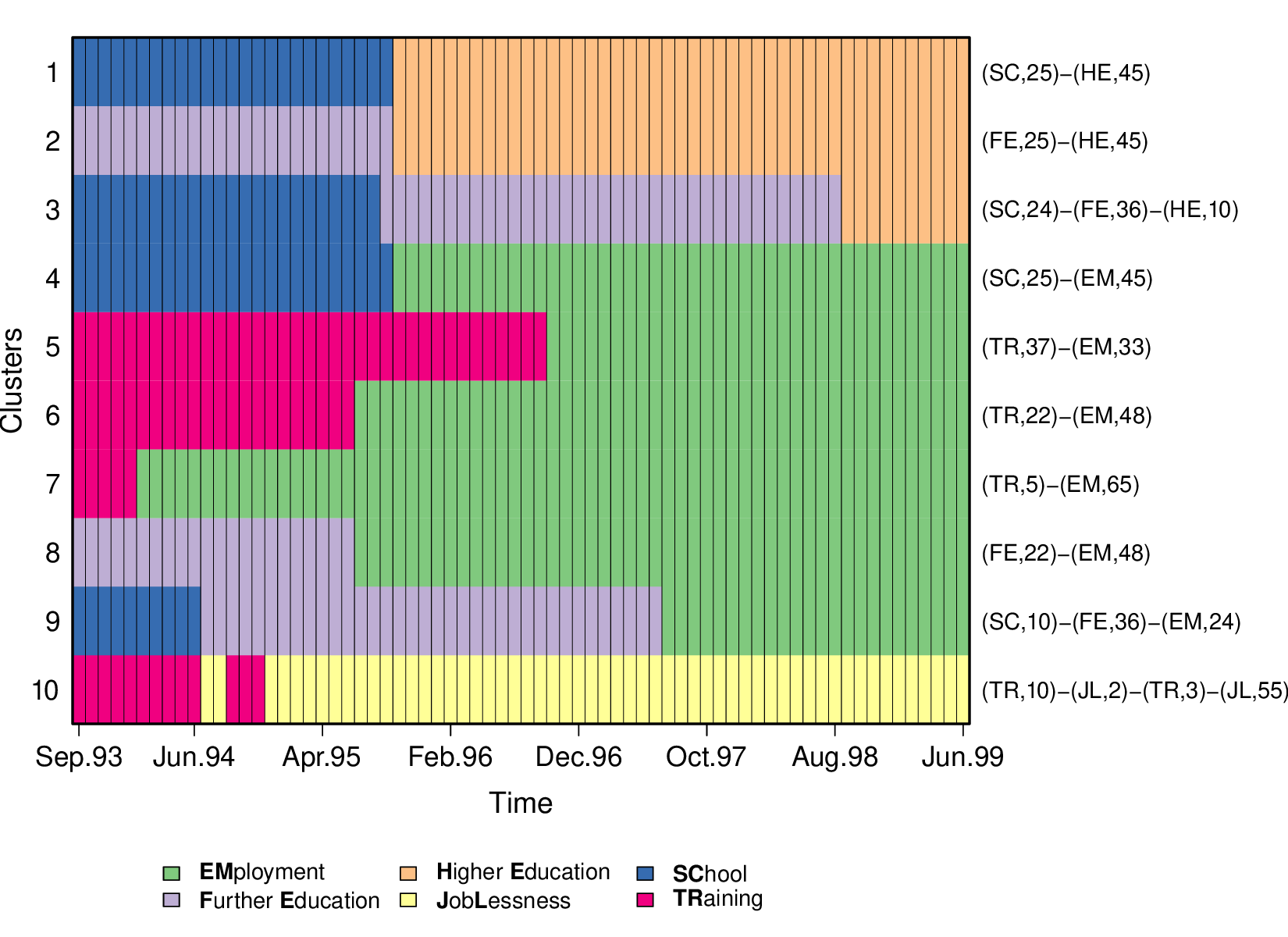}
	\caption{Central sequences of the optimal $G=11$ \textsf{UUN} model with the GCSE5eq gating covariate. The SPS labels on the right characterise each non-noise cluster by the distinct successive states in $\smash{\widehat{\boldsymbol{\theta}}_g}$, with associated durations (in months).\label{Plot:mvadMean}}
\end{figure}

\begin{figure}[H]
	\centering
	\includegraphics[width=\textwidth,keepaspectratio]{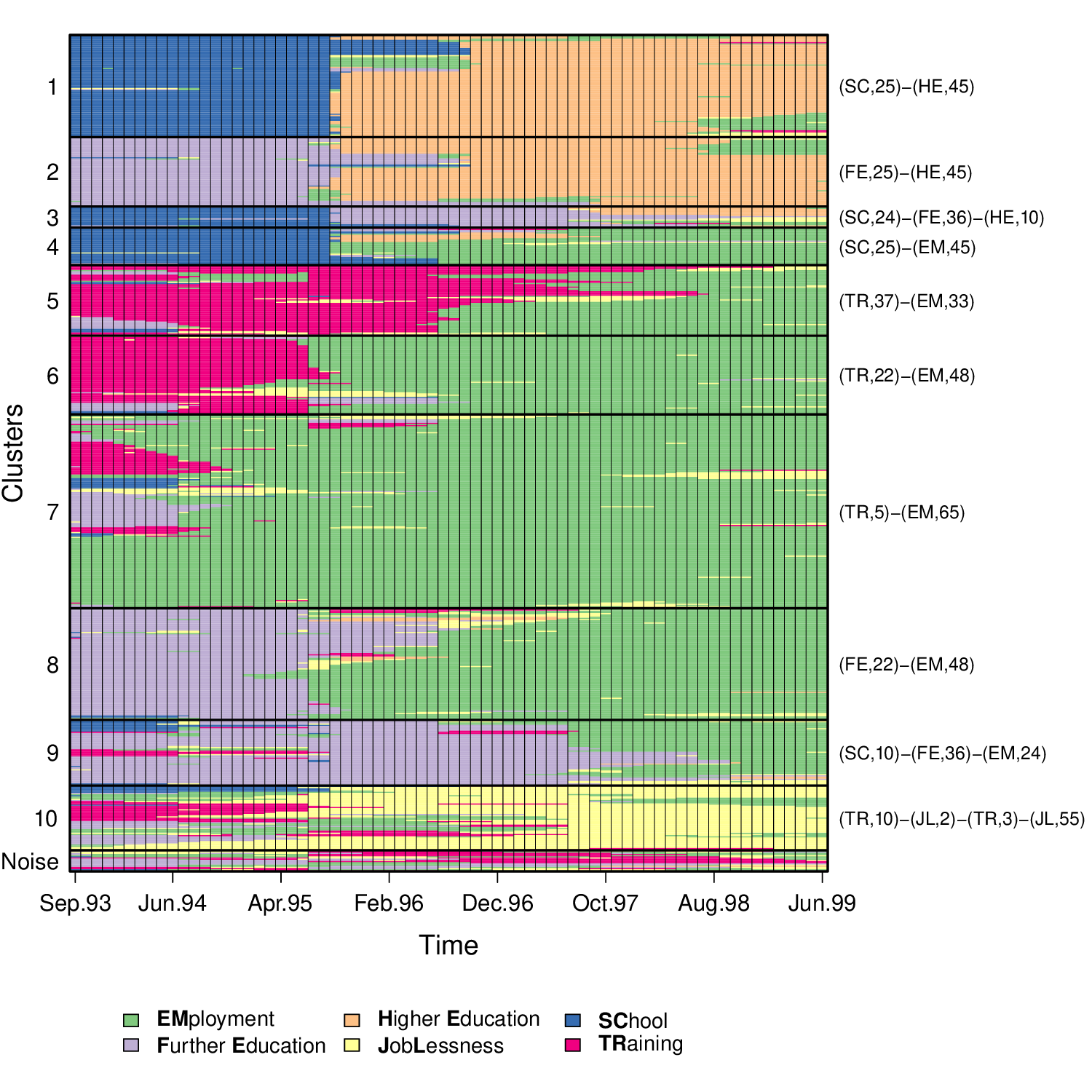}
	\caption{Clusters uncovered under the optimal $G=11$ \textsf{UUN} model with the GCSE5eq gating covariate. The rows correspond to the $n=712$ observed sequences, including duplicate cases previously discarded during model fitting, grouped according to the MAP classification and ordered according to the observed Hamming distance matrix. Each cluster is named according to the SPS representation of $\smash{\widehat{\boldsymbol{\theta}}_g}$.\label{Plot:mvadClusters}}
\end{figure}

To better inform a discussion of the results obtained by the optimal $G=11$ \textsf{UUN} model for the MVAD data, with the covariate GCSE5eq in the NGN gating network, its estimated central sequences are first shown in Figure \ref{Plot:mvadMean}. Seriation has been applied, using the observed Hamming distance matrix and the travelling salesperson combinatorial optimisation algorithm \citep{Hahsler2008}, in order to give consecutive numbers to clusters with similar estimated (weighted) modal sequences. Each cluster's label is derived from the representation of $\smash{\widehat{\boldsymbol{\theta}}_g}$ in State-Permanence-Sequence format (SPS; \citealp{Aassve2007}). The same ordering and labels are used in all subsequent graphical and tabular displays of results. The uncovered clusters are shown in Figure \ref{Plot:mvadClusters}, to which additional seriation has been applied in order to also group the observations within clusters, for visual clarity. Finally, the average time spent in each state by cluster --- weighted by $w_i$ and the estimated cluster membership probabilities --- is shown in Table \ref{Table:MVADmeantime}, along with the cluster sizes.\vspace{-1em}
\enlargethispage{1.5\baselineskip}
\vspace{-1em}
\begin{table}[H]
	\caption{Average durations (in months) spent in each state by cluster, weighted by $\widehat{z}_{i,g}w_i$, for the optimal $11$-component \textsf{UUN} model, with the SPS labels derived from $\smash{\widehat{\boldsymbol{\theta}}_g}$. Estimated cluster sizes $\smash{\widehat{n}_g}$ correspond to the MAP partition.\label{Table:MVADmeantime}}
	\centering
	\scriptsize
	\setlength{\tabcolsep}{4.375pt}
	\begin{tabular}[pos=center]{c l | c | c c c c c c}
		\specialrule{.1em}{.01em}{.01em} 
		\multicolumn{2}{c|}{\multirow{2}{*}{Cluster:\,$g$~(SPS)}} &  \multirow{2}{*}{$\widehat{n}_g$} & \multirow{2}{*}{\textbf{EM}ployment} & \textbf{F}urther  &  \textbf{H}igher  &   \multirow{2}{*}{\textbf{J}ob\textbf{L}essness} &   \multirow{2}{*}{\textbf{SC}hool} &   \multirow{2}{*}{\textbf{TR}aining}\\
		&&&& \textbf{E}ducation & \textbf{E}ducation & & &\\
		\hline\hline		
		1&(SC,25)-(HE,45)               &      $87$&  $3.77$&  $0.29$& $38.45$&  $0.89$& $26.07$&  $0.54$\\
		2&(FE,25)-(HE,45)               &      $59$&  $4.65$& $26.51$& $37.63$&  $0.45$&  $0.76$&  $0.00$\\
		3&(SC,24)-(FE,36)-(HE,10)       &      $18$&  $3.40$& $30.58$& $8.56$&  $4.07$& $21.84$&  $1.56$\\
		4&(SC,25)-(EM,45)               &      $32$& $35.60$&  $1.68$&  $3.63$&  $2.85$& $25.60$&  $0.63$\\
		5&(TR,37)-(EM,33)               &      $60$& $28.29$&  $1.24$&  $0.00$&  $3.38$&  $1.35$& $35.74$\\
		6&(TR,22)-(EM,48)               &      $67$& $45.84$&  $1.47$&  $0.00$&  $3.15$&  $1.51$& $18.03$\\
		7&(TR,5)-(EM,65)                &     $165$& $57.50$&  $2.11$&  $0.00$&  $5.16$&  $1.62$&  $3.62$\\
		8&(FE,22)-(EM,48)               &      $95$& $41.30$& $22.65$&  $0.99$&  $3.04$&  $1.30$&  $0.72$\\
		9&(SC,10)-(FE,36)-(EM,24)       &      $56$& $21.82$& $35.19$&  $0.27$&  $3.99$&  $6.15$&  $2.58$\\
		10&(TR,10)-(JL,2)-(TR,3)-(JL,55)&      $55$& $8.40$&  $3.38$&  $0.22$& $42.89$&  $4.20$&  $10.91$\\
		Noise&---                       &      $18$& $21.50$& $11.42$&  $1.19$& $14.51$&  $2.20$& $19.18$\\
		\specialrule{.1em}{.01em}{.01em} 
	\end{tabular}
\end{table}
This solution tends to group individuals who experience trajectories that are similar or that differ only for relatively short periods. In particular, the dominating combinations of states experienced over time are clearly identified, and differences in durations and/or age at transition are quite limited in size. Within clusters, substantial reduction of misalignments and/or differences in the durations of spells are evident. Ultimately, the partition~is characterised not only by the sequencing (i.e. the experienced, ordered combinations of~states), but also by the spell durations and the ages at transitions which appear mostly homogeneous within clusters. This can be explained by the fact that cases in the identified groups tend to dedicate the same period of time (spells of 1, 2, or 3 years) to further/higher education and/or training. This is interesting because one might expect the chosen dissimilarity metric, as it based on the Hamming distance, to attach higher importance to the sequencing.

The $11$-cluster solution for the MVAD data separates individuals who continued in school (clusters 1, 3, and 4), or otherwise prolonged their studies after the end of compulsory education (clusters 2, 8, and 9), from those who entered the labour market (clusters 5, 6, and 7). The clear division visible for some clusters in Figure \ref{Plot:mvadClusters} around Autumn 1995, when new semesters of further and higher education commenced and the majority of those still remaining in school had eventually left, corresponds to the time point in Figure \ref{Plot:MVADEntropy} after which the entropies declined. Interestingly, individuals who experienced prolonged periods of unemployment are mostly isolated in cluster 10; this is particularly important because the Status Zero Survey aimed to identify such `at risk' subjects. From this we conclude that youth unemployment in Northern Ireland in this period was predominantly a problem of small numbers experiencing long spells of non-participation in the labour market rather than large numbers dipping into brief, frictional spells.

Clusters 1, 3, and 4 include subjects who continued school for about two years, presumably to retake previously failed examinations or to pursue academic or vocational qualifications. These individuals are split into two groups depending on whether they continued their studies (FE: cluster 3, or HE: cluster 1) or were employed directly (cluster 4). Clusters 2, 8, and 9 group subjects who initially entered further education, for about two years (clusters 2 and 8) or more (cluster 9). Most subjects in clusters 8 and 9 entered employment directly after further education, whereas the vast majority of those in cluster 2 transitioned to higher education, where they remained until the end of the observation period.

As for the clusters of individuals who moved quickly to the labour market after the end of compulsory education, it is possible to distinguish between individuals who almost immediately found a job and remained in employment for most of the observation period (the large cluster 7) and individuals who entered government-supported training schemes (clusters 5 and 6). A further separation is between subjects who were employed after about 2 years of training (cluster 6) and those who participated in training for a much longer period (cluster 5). Importantly, most of the individuals in these two clusters were able to find a job even if some respondents experienced some periods of unemployment.  

It is interesting to observe that the cluster of careers dominated by persistent unemployment (cluster 10) is characterised by different experiences at the end of the compulsory education period. Indeed, some subjects entered employment directly after the end of compulsory education but left or lost their job after some months, while some prolonged their education before becoming unemployed. However, the majority entered a training period that did not evolve into steady employment.

Notably, the optimal model identified is a \textsf{UUN} model, i.e. one whose precision parameters vary across both clusters and time points. Thus, model selection favours a flexible, heavily-parameterised MEDseq variant which, while based on the simple Hamming distance, has cluster-specific and period-specific costs which allow element-wise mismatches between sequences and the central sequences in different time periods in different clusters to contribute differently to the overall distance measure. While a display of the estimated precision parameters is omitted, for brevity, their values can be easily examined via the \texttt{MEDseq} \textsf{R} package. Nonetheless, it is already clear that the model captures different degrees of heterogeneity in the cluster-specific state distributions of each month.

The coefficients of the gating network with associated WLBS standard errors are given in Table \ref{Table:mvadGating}, from which a number of interesting effects can be identified. The interpretation of the effects of the covariates is made clearer by virtue of there being just one included after stepwise selection. For completeness, gating network coefficients and associated WLBS standard errors for the model with all covariates included are provided in Appendix \ref{Section:AllGating}.
\begin{table}[H]
	\caption{Multinomial logistic regression coefficients and associated WLBS standard errors (in parentheses), with SPS labels, for the NGN gating network of the optimal $G=11$ \textsf{UUN} model with the GCSE5eq~covariate. Recall that GCSE5eq=$1$ for subjects who achieved 5 or more grades at A--C (or equivalent) in GCSE exams.\label{Table:mvadGating}}
	\centering
	\scriptsize
	\extrarowheight 2.5pt
	\setlength{\tabcolsep}{5pt}
	
	\begin{tabular}[pos=center]{c l | r r r r}
		\specialrule{.1em}{.01em}{.01em} 
		\multicolumn{2}{c|}{Cluster:\,$g$~(SPS)}  & \multicolumn{2}{c}{(Intercept)} & \multicolumn{2}{c}{GCSE5eq}\\
		\hline\hline
		$1$&(SC,25)-(HE,45)               & \multicolumn{2}{c}{---}&\multicolumn{2}{c}{---}\\
		$2$&(FE,25)-(HE,45)               &       $-0.95$& ($0.44$)&   $-0.47$& ($0.49$)\\
		$3$&(SC,24)-(FE,36)-(HE,10)       &       $-0.46$& ($0.63$)&   $-1.23$& ($0.73$)\\
		$4$&(SC,25)-(EM,45)               &        $0.58$& ($0.44$)&   $-2.18$& ($0.58$)\\
		$5$&(TR,37)-(EM,33)               &        $1.03$& ($0.38$)&   $-3.43$& ($0.55$)\\
		$6$&(TR,22)-(EM,48)               &        $1.19$& ($0.35$)&   $-3.73$& ($0.50$)\\
		$7$&(TR,5)-(EM,65)                &        $1.70$& ($0.32$)&   $-4.09$& ($0.47$)\\
		$8$&(FE,22)-(EM,48)               &        $0.60$& ($0.38$)&   $-2.20$& ($0.42$)\\
		$9$&(SC,10)-(FE,36)-(EM,24)       &        $0.95$& ($0.39$)&   $-3.20$& ($0.55$)\\
		$10$&(TR,10)-(JL,2)-(TR,3)-(JL,55)&        $0.90$& ($0.36$)&   $-3.73$& ($0.72$)\\
		\specialrule{.1em}{.01em}{.01em} 
	\end{tabular}
\end{table}
Relative to the reference cluster (cluster 1), characterised by those who prolonged their schooling for two years to sit A-level exams before successfully transitioning to higher education, all slope coefficients are notably negative. All students achieving 5 or more grades at A--C in GCSE exams are therefore less likely to belong to all other clusters, relative to cluster 1. Thus, the reference level for the effect of GCSE5eq is appropriate and the interpretation is guided only by the magnitude of the slope coefficients and their associated standard errors, as well as the intercepts. Firstly, the effects for cluster 2 and 3, capturing other subjects who were in higher education by the end of the observation period, appear slight (on the basis of the size of the standard errors of their slopes). Coupled with the negative intercepts for these clusters, this suggests, as expected, that more academically inclined students tend to prolong their education in order to improve their job prospects.

Conversely, all other intercepts are positive and all other slope coefficients appear to be significantly different from $0$. We can say, therefore, despite the two-year continuation in school of subjects in cluster 4, that students who do well in GCSE exams are less likely to belong to this cluster. Furthermore, we can see the coefficient magnitudes increasing and the standard errors decreasing as we move from cluster 5 to cluster 7. As these clusters are distinguished only by the length of the training period prior to securing stable employment, this again suggests that academically poor students are quick to find a job, presumably in an unskilled capacity. Similar conclusions can be drawn for clusters 8 and 9, i.e. subjects who secured employment of some kind after some time in further education rather than third-level education. Finally, those who achieved 5 or more high GCSE grades are less likely to experience persistent spells of joblessness (cluster 10).

\begin{figure}[H]
	\centering
	\includegraphics[width=0.7125\textwidth,keepaspectratio]{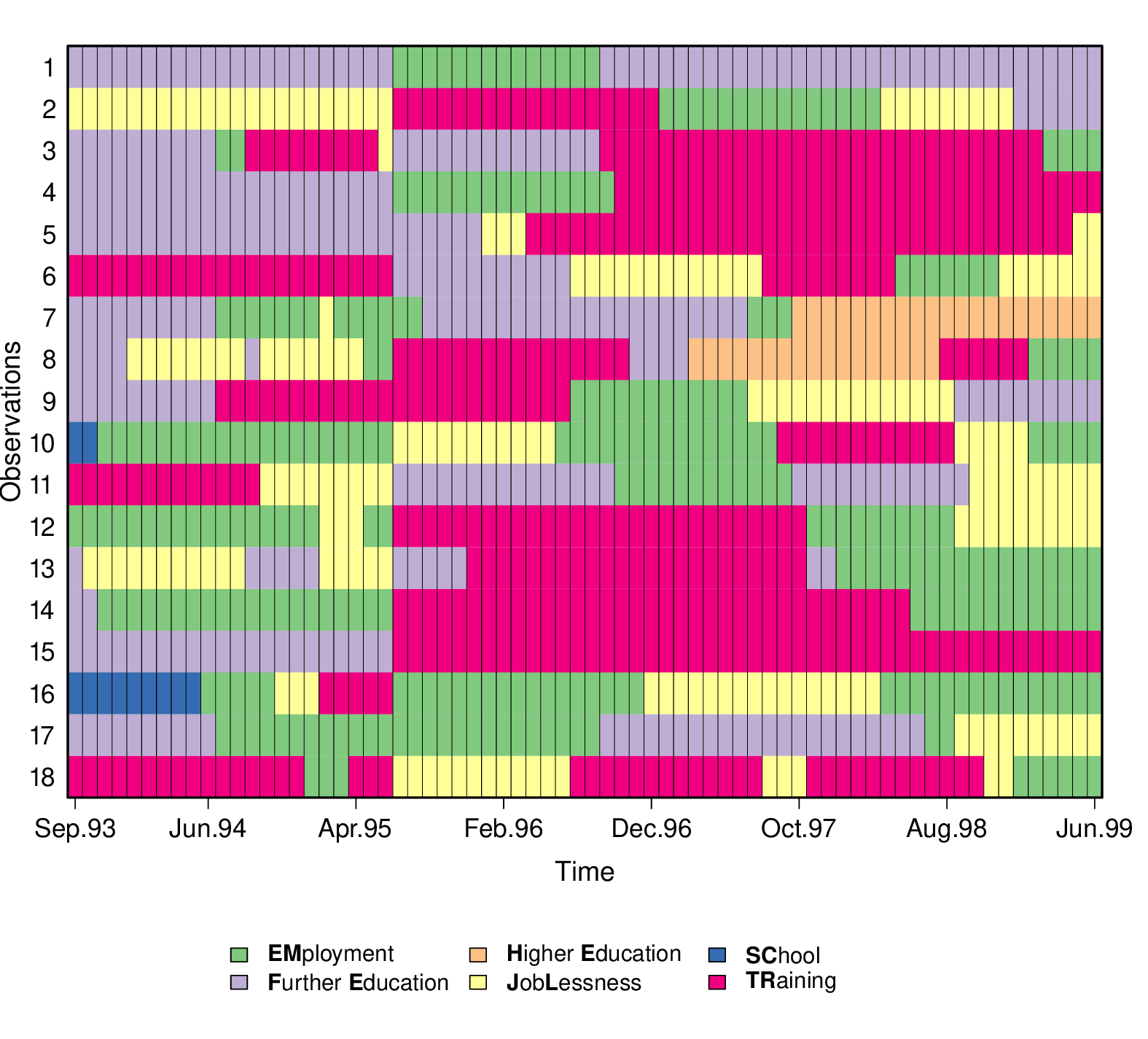}
	\vspace*{-1mm}
	\caption{Observations assigned to the noise component of the optimal $G=11$ \textsf{UUN} model with the GCSE5eq covariate in the NGN gating network.}
	\label{Plot:MVADNoise}
\end{figure}

The optimal $G=11$ \textsf{UUN} model contains a uniform noise component. The BIC chooses such a model over $G=10$ models without a noise component and $G=11$ models with all non-noise components. Detecting outliers in this way allows the remaining non-noise clusters to be modelled more clearly. Figure \ref{Plot:MVADNoise} focuses on the noise component, which isolates errant, directionless subjects who don't neatly fit into any of the defined clusters and transition quite frequently between states. This includes transitions in and out of further education, employment, and training. Most subjects here are early school-leavers. Under the model's NGN gating network, the probability of belonging to this noise component is constant ($\approx 0.025$) and independent of the included GCSE5eq covariate.

\section[Conclusion]{Conclusion}
\label{Section:Conclusion}

The Status Zero Survey followed a sample of Northern Irish youths over a six-year period, recording their employment activities at monthly intervals, in order to explore their unfolding career trajectories and identify those at risk of prolonged unemployment. Here we present a model-based clustering approach, with the aims of assessing how many typical trajectories there are, what kinds of typical trajectories there are, and what kinds of individuals are more likely to experience which kinds of trajectories. Our approach is contrasted to heuristic approaches previously employed in analyses of these data. In \citet{mvad2002}, Ward's hierarchical clustering algorithm is applied to an OM dissimilarity matrix to identify relevant patterns in the data, with subjective costs. Notably, reference is not made to the associated covariates until the uncovered clustering structure is examined. In particular, MLR is used to relate the hard assignments of the sequences to clusters to a set of baseline covariates. It is also notable that the sampling weights are incorporated only in the MLR stage and not in the clustering itself. This is arguably a three-step approach, comprising the computation of pairwise string distances using OM (or some other distance metric), the hierarchical or partition-based clustering, and the (weighted) MLR.

MEDseq models, conversely, offer a more coherent `one-step' model-based approach.~The sequences are modelled directly using a finite mixture of exponential-distance models, with the Hamming distance and weighted variants thereof employed as the distance metric. A range of precision parameter settings have been explored to allow different time points contribute differently to the overall distance. Thus, varying degrees of parsimony are accommodated. Sampling weights are accounted for by weighting each observation's contribution to the pseudo likelihood. Dependency on covariates is introduced by relating the cluster membership probabilities to covariates under the mixture of experts framework. Thus, MEDseq models treat the weights, the relation of covariates to clusters, and the clustering itself simultaneously. Hence, MEDseq provides a coherent framework for estimating the number of clusters, identifying the relevant features of these patterns, and assessing whether these patterns are somehow influenced or shaped by the subjects' background characteristics.

Model selection in the MEDseq setting identifies a reasonable solution for the MVAD data and shows that clustering the sequences in a holistic manner allows new insights to be gleaned from these data. In particular, $11$ distinct components are found, of which $10$ have interpretable typical trajectories and one is an additional noise component which captures deviant cases. Thus, supported by the use of an information criterion appropriate for this model-based analysis, a more granular view of the MVAD cohort than the $5$ groups uncovered in \citet{mvad2002} is provided. Furthermore, allowing for the other covariates with which the sampling weights used here are defined, GCSE exam performance at the end of the compulsory education period is found to be the most single most important predictor of cluster membership.

Opportunities for future research are varied and plentiful. Co-clustering approaches could be used to simultaneously provide clusters of the observed sequence trajectories and the time periods \citep{Govaert2013}. Such an approach could be especially useful for the \textsf{UUN} model type identified as optimal for the MVAD data, as it would reduce the number of within-cluster period-specific precision parameters required. Indeed, parsimony has been achieved in a similar fashion in the context of finite mixtures with Markov components \citep{Melnykov2016}. Additionally, grouping trajectories across time as well would enable more efficient summaries of the durations of the spells in specific states, which tend to be long for the MVAD data. In particular, using co-clustering approaches which respect the ordering of the sequences by restricting the column-wise clusters to form contingent blocks would be particularly desirable. Indeed, failure to fully account for the temporal ordering of events, due to the invariance of the Hamming distance to permutations of the time periods, is a general limitation of our framework which future work will endeavour to address.

It may also be of interest for other applications to extend the MEDseq models to accommodate sequences of different lengths, for which the Hamming distance is not defined. These different lengths could be attributable to missing data, either by virtue of sequences not starting on the same date, shorter follow-up time for some subjects, or non-response for some time points. While the Hamming distance is only defined for equal-length strings, adapting the MEDseq models to such a setting would be greatly simplified if aligning the sequences of different lengths is straightforward. Another limitation of MEDseq models is that time-varying covariates are not accommodated in the gating network. Notably, neither of these concerns are relevant for the MVAD data.\enlargethispage{\baselineskip}

However, our analysis of the MVAD data is limited by two aspects of the gating network portion of our framework. The first substantive limitation relates to the WLBS approach used for quantifying uncertainty in the MLR coefficients. As the sampling weights arise from stratification, the standard errors obtained in this fashion are approximate. Thus, examining alternative approaches to produce fully valid variance estimates in the MEDseq setting in the presence of complex sampling designs is an interesting future research avenue. 

The second limitation relates to the stepwise procedure used to identify relevant covariates. As this strategy depends on an information criterion, namely the BIC, whose penalty term is based on a parameter count, it may be prudent to relax the assumption that gating covariates must affect all components. As the number of components chosen~here ($G=11$) is moderately large, a large number of extra parameters are associated with each extra covariate (see Appendix \ref{Section:ParameterCount}). Thus, only GCSE5eq is identified as optimal,~as~it is significantly associated with many of the typical trajectories. However, we note, for~\mbox{example}, that Catholics are largely underrepresented in cluster 7 and largely overrepresented in~\mbox{cluster} 10 (characterised by persistent employment and persistent joblessness, respectively) despite the omission of the covariate indicating religious affiliation from the optimal model. Incorporating regularisation penalties into the MLR to shrink certain gating network coefficients to zero could thus be a fruitful alternative to the present stepwise covariate-selection method.

Another potential extension is to consider MEDseq models with alternative distance metrics. The distance metric in \citet{Garcia2015}, which accounts for the temporal correlation in categorical sequences, is of particular interest; so, too, is OM. In general, heuristic distance-based clustering (including fuzzy methods) can more easily accommodate more sophisticated distances, while changing the MEDseq distance metric fundamentally alters the model, which needs the normalising constant and the conditional maximisation steps for parameter estimation to be tailored to the choice of metric. 

MEDseq models, by virtue of being based on the Hamming distance for computational reasons, implicitly assume substitution-cost matrices with zero along the diagonal and a single value common to all other entries. The relationship between the exponent of an EDM based on the Hamming distance and the Hamming distance itself (with a common cost, typically equal to $1$) is apparent from the fact that multiplying the substitution-cost matrix by any positive scalar, as per normalised variants of the Hamming distance \citep{Elzinga2007, TraMineR2011}, yields the same model, because its value is absorbed into~$\lambda$. This is also the case for models employing weighted Hamming distance variants under which the precision parameters, and hence the \emph{otherwise} common substitution costs, vary across clusters and/or time points. However, all model types in the MEDseq family cannot account for situations in which some states are more different than others --- e.g. one where the cost associated with moving from employment to joblessness is assumed to be greater than the cost associated with moving from school to training --- as they assume that substitution costs are the same between each pair of states. Such concerns are most pronounced when there is an explicit ordering to the states, e.g. education levels \citep{Studer2016}.

Basing MEDseq on OM, for instance, would require the subjective specification, or preferably estimation, of the $v\negthinspace\left(v-1\right)\negthinspace/2$ off-diagonal entries of symmetric substitution-cost matrices. Potentially, as per the range of precision settings used for the MVAD application, the substitution-cost matrices could also be allowed to vary across clusters and/or time points. However, the normalising constant under an EDM using OM depends both on the heterogeneous substitution costs and on $\boldsymbol{\theta}$ and is unavailable in closed form, thereby greatly complicating model fitting. Indeed, dependence on $\boldsymbol{\theta}$ renders even offline pre-computation of the normalising constant infeasible for even moderately large $T$ or $v$. Truncation of the sum over all sequences or importance sampling techniques could be used to address the intract\-ability. Though not a concern for the MVAD data, as one substitution is equivalent to a deletion and an insertion for equal-length sequences, considering insertions and deletions also would present further challenges. In any case, some level of approximation would be required, while the ECM algorithm for MEDseq models based on simple matching is exact.

As well as removing the normalising constant's dependence on $\boldsymbol{\theta}$, another positive consequence of the homogeneity of substitution costs with respect to pairs of states under the Hamming distance is that the ECM algorithm used for parameter estimation scales well with the sequence length $T$ and the size of the alphabet $v$, especially since such normalising constants need to be computed once, $G$ times, or $G-1$ times \emph{per iteration}, depending on the precision parameter settings. Though potentially restrictive, having only one parameter associated with each substitution-cost matrix, regardless of its order $v$, helps address concerns about overparameterisation \citep{Studer2016}, especially when the substitution costs implied by the precision parameter(s) vary across clusters and/or time points. 

Furthermore, it is likely that results on the MVAD data would not differ greatly with OM used in place of the Hamming distance, particularly for models where $\lambda$ varies across clusters and/or time points, save for a solution with potentially fewer clusters being found. Indeed, \citet{mvad2002} also consider a setting with \emph{common} substitution costs and find that their results do not greatly differ from their solution with state-dependent costs. This implies that the notion that some states in the MVAD data are closer to each other than others can be questioned. Ultimately, the \textsf{UUN} model adopted here preserves the timing of events, by prohibiting time-warping insertion and deletion operations, while accounting (in a cluster-specific fashion) for the timing, as well as the number, of element-wise mismatches between sequences, in such a way that all states are assumed to~be equally different. Given the correspondence between Hamming distance weights, precision param\-eters, and implicit substitution costs in MEDseq models, it is notable that these are treated as parameters rather than inputs, and are thus estimated as part of model fitting rather than pre-specified along with the matrix of pairwise distances between sequences.

Overall, our analysis of the MVAD data provides a more granular view of the cohort of Northern Irish youths than previously available, supplemented by interpretable parameter estimates achieved through a coherent model-based framework. The MEDseq model family appears promising from the perspective of reconciling the distance-based and model-based cultures within the SA community. Indeed, the results for the MVAD data are encouraging in this respect; they seem to suggest that the unconstrained precision parameter settings adequately address the misalignment issues inherent in the use of the Hamming distance. It remains to be seen if this holds for more turbulent sequences, e.g. those related to employment activities tracked over longer periods.

\subsection*{Acknowledgements}
\small
This publication has emanated from research conducted with the financial support of Science \mbox{Foundation} Ireland under Grant number SFI/12/RC/2289\_P2. Additionally, R. Piccarreta\linebreak acknowledges the support from MIUR-PRIN 2017 project 20177BRJXS. For the purpose of Open Access, the authors have applied a CC BY public copyright licence to any Author Accepted Manuscript version arising from this submission. The authors also thank Matthias Studer and members of the Sequence Analysis Association for helpful discussions.
\normalsize

\clearpage
\bibliographystyle{Chicago}\interlinepenalty=10000
\bibliography{MEDseq_JRSSA_bib}

\begin{thebibliography}{}

\bibitem[\protect\citeauthoryear{Aassve, Billari, and Piccarreta}{Aassve
  et~al.}{2007}]{Aassve2007}
Aassve, A., F.~C. Billari, and R.~Piccarreta (2007).
\newblock Strings of adulthood: a sequence analysis of young {B}ritish women's
  weekly work-family trajectories.
\newblock {\em European Journal of Population\/}~{\em 23\/}(3), 369--388.

\bibitem[\protect\citeauthoryear{Abbott and Forrest}{Abbott and
  Forrest}{1986}]{Abbott1986}
Abbott, A. and J.~Forrest (1986).
\newblock {O}ptimal matching methods for historical sequences.
\newblock {\em Journal of Interdisciplinary History\/}~{\em 16\/}(3), 471--494.

\bibitem[\protect\citeauthoryear{Abbott and Hrycak}{Abbott and
  Hrycak}{1990}]{Abbott1990}
Abbott, A. and A.~Hrycak (1990).
\newblock Measuring resemblance in sequence data: an optimal matching analysis
  of musician's careers.
\newblock {\em American Journal of Sociology\/}~{\em 96\/}(1), 145--185.

\bibitem[\protect\citeauthoryear{Agresti}{Agresti}{2002}]{Agresti2002}
Agresti, A. (2002).
\newblock {\em {Categorical Data Analysis}}.
\newblock New York: John Wiley \& Sons.

\bibitem[\protect\citeauthoryear{Airoldi, Blei, Erosheva, and Fienberg}{Airoldi
  et~al.}{2014}]{Airoldi2014}
Airoldi, E.~M., D.~M. Blei, E.~A. Erosheva, and S.~E. Fienberg (2014).
\newblock {\em {Handbook of Mixed Membership Models and Their Applications}}.
\newblock New York, USA: Chapman and Hall/CRC Press.

\bibitem[\protect\citeauthoryear{Armstrong, Istance, Loudon, McCready, Rees,
  and Wilson}{Armstrong et~al.}{1997}]{Armstrong1997}
Armstrong, D., D.~Istance, R.~Loudon, S.~McCready, G.~Rees, and D.~Wilson
  (1997).
\newblock `{S}tatus 0': a socio-economic study of young people on the margin.
\newblock {Belfast: Training and Employment Agency}, {Northern Ireland Economic
  Research Centre}.

\bibitem[\protect\citeauthoryear{Bakk and Kuha}{Bakk and Kuha}{2018}]{Bakk2018}
Bakk, Z. and J.~Kuha (2018).
\newblock Two-step estimation of models between latent classes and external
  variables.
\newblock {\em Psychometrika\/}~{\em 83\/}(4), 871--892.

\bibitem[\protect\citeauthoryear{Banerjee, Dhillon, Ghosh, and Sra}{Banerjee
  et~al.}{2005}]{Banerjee2005}
Banerjee, A., I.~S. Dhillon, J.~Ghosh, and S.~Sra (2005).
\newblock Clustering on the unit hypersphere using von {M}ises-{F}isher
  distributions.
\newblock {\em Journal of Machine Learning Research\/}~{\em 6\/}(46),
  1345--1382.

\bibitem[\protect\citeauthoryear{Banfield and Raftery}{Banfield and
  Raftery}{1993}]{Banfield1993}
Banfield, J. and A.~E. Raftery (1993).
\newblock {M}odel-based {G}aussian and non-{G}aussian clustering.
\newblock {\em Biometrics\/}~{\em 49\/}(3), 803--821.

\bibitem[\protect\citeauthoryear{Billari}{Billari}{2001}]{Billari2001}
Billari, F.~C. (2001).
\newblock The analysis of early life courses: complex description of the
  transition to adulthood.
\newblock {\em Journal of Population Research\/}~{\em 18\/}(2), 119--142.

\bibitem[\protect\citeauthoryear{Bishop}{Bishop}{2006}]{Bishop2006}
Bishop, C.~M. (2006).
\newblock {\em {Pattern Recognition and Machine Learning}}.
\newblock New York: Springer.

\bibitem[\protect\citeauthoryear{B{\"{o}}hning, Dietz, Schaub, Schlattmann, and
  Lindsay}{B{\"{o}}hning et~al.}{1994}]{Boehning1994}
B{\"{o}}hning, D., E.~Dietz, R.~Schaub, P.~Schlattmann, and B.~G. Lindsay
  (1994).
\newblock The distribution of the likelihood ratio for mixtures of densities
  from the one-parameter exponential family.
\newblock {\em Annals of the Institute of Statistical Mathematics\/}~{\em
  46\/}(2), 373--388.

\bibitem[\protect\citeauthoryear{Bouveyron, Celeux, Murphy, and
  Raftery}{Bouveyron et~al.}{2019}]{Bouveyron2019}
Bouveyron, C., G.~Celeux, T.~B. Murphy, and A.~E. Raftery (2019).
\newblock {\em {Model-Based Clustering and Classification for Data Science:
  With Applications in R}}.
\newblock Cambridge Series in Statistical and Probabilistic Mathematics.
  Cambridge: Cambridge University Press.

\bibitem[\protect\citeauthoryear{Celeux and Govaert}{Celeux and
  Govaert}{1992}]{Celeux1992}
Celeux, G. and G.~Govaert (1992).
\newblock A classification {EM} algorithm for clustering and two stochastic
  versions.
\newblock {\em Computational Statistics \& Data Analysis\/}~{\em 14\/}(3),
  315--332.

\bibitem[\protect\citeauthoryear{Chambers and Skinner}{Chambers and
  Skinner}{2003}]{Chambers2003}
Chambers, R.~L. and C.~J. Skinner (2003).
\newblock {\em {Analysis of Survey Data}}.
\newblock Chichester: John Wiley \& Sons.

\bibitem[\protect\citeauthoryear{Dayton and Macready}{Dayton and
  Macready}{1988}]{Dayton1988}
Dayton, C.~M. and G.~B. Macready (1988).
\newblock {C}oncomitant-variable latent-class models.
\newblock {\em Journal of the American Statistical Association\/}~{\em
  83\/}(401), 173--178.

\bibitem[\protect\citeauthoryear{de~Amorim}{de~Amorim}{2015}]{deAmorim2015}
de~Amorim, R.~C. (2015).
\newblock Feature relevance in {W}ard's hierarchical clustering using the
  $\mathrm{L_p}$ norm.
\newblock {\em Journal of Classification\/}~{\em 32\/}(1), 46--62.

\bibitem[\protect\citeauthoryear{Dempster, Laird, and Rubin}{Dempster
  et~al.}{1977}]{Dempster1977}
Dempster, A.~P., N.~M. Laird, and D.~B. Rubin (1977).
\newblock {M}aximum likelihood from incomplete data via the {EM} algorithm.
\newblock {\em Journal of the Royal Statistical Society: Series B (Statistical
  Methodology)\/}~{\em 39\/}(1), 1--38.

\bibitem[\protect\citeauthoryear{Diaconis}{Diaconis}{1988}]{Diaconis1988}
Diaconis, P. (1988).
\newblock {\em Group Representations in Probability and Statistics}, Volume~11
  of {\em Lecture Notes --- Monograph Series}.
\newblock Hayward, CA, USA: Institute of Mathematical Statistics.

\bibitem[\protect\citeauthoryear{D'Urso}{D'Urso}{2016}]{Durso2016}
D'Urso, P. (2016).
\newblock Fuzzy clustering.
\newblock In C.~Hennig, M.~Meila, F.~Murtagh, and R.~Rocci (Eds.), {\em
  Handbook of Cluster Analysis}, Chapter~24, pp.\  245--575. New York: Chapman
  and Hall.

\bibitem[\protect\citeauthoryear{D'Urso and Massari}{D'Urso and
  Massari}{2013}]{Durso2013}
D'Urso, P. and R.~Massari (2013).
\newblock Fuzzy clustering of human activity patterns.
\newblock {\em Fuzzy Sets and Systems\/}~{\em 215}, 29--54.

\bibitem[\protect\citeauthoryear{Elzinga}{Elzinga}{2007}]{Elzinga2007}
Elzinga, C.~H. (2007).
\newblock {Sequence analysis: metric representations of categorical time
  series}.
\newblock Technical report, Department of Social Science Research Methods,
  Vrije Universiteit, Amsterdam.

\bibitem[\protect\citeauthoryear{Fligner and Verducci}{Fligner and
  Verducci}{1986}]{Fligner1986}
Fligner, M.~A. and J.~S. Verducci (1986).
\newblock Distance based ranking models.
\newblock {\em Journal of the Royal Statistical Society: Series B (Statistical
  Methodology)\/}~{\em 48\/}(3), 359--369.

\bibitem[\protect\citeauthoryear{Gabadinho, Ritschard, M{\"{u}}ller, and
  Studer}{Gabadinho et~al.}{2011}]{TraMineR2011}
Gabadinho, A., G.~Ritschard, N.~S. M{\"{u}}ller, and M.~Studer (2011).
\newblock {A}nalyzing and visualizing state sequences in \textsf{R} with
  {T}ra{M}ine{R}.
\newblock {\em Journal of Statistical Software\/}~{\em 40\/}(4), 1--37.

\bibitem[\protect\citeauthoryear{Garc{\'{i}}a-Magari{\~{n}}os and
  Vilar}{Garc{\'{i}}a-Magari{\~{n}}os and Vilar}{2015}]{Garcia2015}
Garc{\'{i}}a-Magari{\~{n}}os, M. and J.~A. Vilar (2015).
\newblock A framework for dissimilarity-based partitioning clustering of
  categorical time series.
\newblock {\em Data Mining and Knowledge Discovery\/}~{\em 29\/}(2), 466--502.

\bibitem[\protect\citeauthoryear{Gormley and Fr{\"{u}}hwirth-Schnatter}{Gormley
  and Fr{\"{u}}hwirth-Schnatter}{2019}]{GormleySchnatter2019}
Gormley, I.~C. and S.~Fr{\"{u}}hwirth-Schnatter (2019).
\newblock Mixtures of experts models.
\newblock In S.~Fr{\"{u}}hwirth-Schnatter, G.~Celeux, and C.~P. Robert (Eds.),
  {\em {Handbook of Mixture Analysis}}, Chapter~12, pp.\  279--316. London:
  Chapman and Hall/CRC Press.

\bibitem[\protect\citeauthoryear{Govaert and Nadif}{Govaert and
  Nadif}{2013}]{Govaert2013}
Govaert, G. and M.~Nadif (2013).
\newblock {\em {Co-Clustering: Models, Algorithms and Applications}}.
\newblock London: ISTE-Wiley.

\bibitem[\protect\citeauthoryear{Gower}{Gower}{1971}]{Gower1971}
Gower, J.~C. (1971).
\newblock A general coefficient of similarity and some of its properties.
\newblock {\em Biometrics\/}~{\em 27\/}(4), 857--871.

\bibitem[\protect\citeauthoryear{Hahsler, Hornik, and Buchta}{Hahsler
  et~al.}{2008}]{Hahsler2008}
Hahsler, M., K.~Hornik, and C.~Buchta (2008).
\newblock Getting things in order: an introduction to the \textsf{R} package
  seriation.
\newblock {\em Journal of Statistical Software\/}~{\em 25\/}(3), 1--34.

\bibitem[\protect\citeauthoryear{Hamming}{Hamming}{1950}]{Hamming1950}
Hamming, R.~W. (1950).
\newblock {E}rror detecting and error correcting codes.
\newblock {\em The Bell System Technical Journal\/}~{\em 29\/}(2), 147--160.

\bibitem[\protect\citeauthoryear{Helske and Helske}{Helske and
  Helske}{2019}]{seqHMM2019}
Helske, S. and J.~Helske (2019).
\newblock {M}ixture hidden {M}arkov models for sequence data: the {seqHMM}
  package in \textsf{R}.
\newblock {\em Journal of Statistical Software\/}~{\em 88\/}(3), 1--32.

\bibitem[\protect\citeauthoryear{Helske, Helske, and Eerola}{Helske
  et~al.}{2016}]{Helske2016}
Helske, S., J.~Helske, and M.~Eerola (2016).
\newblock {Analysing complex life sequence data with hidden Markov modeling}.
\newblock In G.~Ritschard and M.~Studer (Eds.), {\em {LaCOSA II: Proceedings of
  International Conference on Sequence Analysis and Related Methods}}, pp.\
  209--240.

\bibitem[\protect\citeauthoryear{Hoos and St{\"{u}}tzle}{Hoos and
  St{\"{u}}tzle}{2004}]{Hoos2004}
Hoos, H. and T.~St{\"{u}}tzle (2004).
\newblock {\em {Stochastic Local Search: Foundations and Applications}}.
\newblock San Francisco, CA, USA: Morgan Kaufmann Publishers Inc.

\bibitem[\protect\citeauthoryear{Huang}{Huang}{1997}]{Huang1997}
Huang, Z. (1997).
\newblock A fast clustering algorithm to cluster very large categorical data
  sets in data mining.
\newblock In H.~Lu, H.~Motoda, and H.~Luu (Eds.), {\em {KDD: Techniques and
  Applications}}, pp.\  21--34. Singapore: World Scientific.

\bibitem[\protect\citeauthoryear{Irurozki, Calvo, and Lozano}{Irurozki
  et~al.}{2019}]{Irurozki2019}
Irurozki, E., B.~Calvo, and J.~A. Lozano (2019).
\newblock {M}allows and generalized {M}allows model for matchings.
\newblock {\em Bernoulli\/}~{\em 25\/}(2), 1160--1188.

\bibitem[\protect\citeauthoryear{Jacobs, Jordan, Nowlan, and Hinton}{Jacobs
  et~al.}{1991}]{Jacobs1991}
Jacobs, R.~A., M.~I. Jordan, S.~J. Nowlan, and G.~E. Hinton (1991).
\newblock {A}daptive mixtures of local experts.
\newblock {\em Neural Computation\/}~{\em 3\/}(1), 79--87.

\bibitem[\protect\citeauthoryear{Kaufman and Rousseeuw}{Kaufman and
  Rousseeuw}{1990}]{Kaufman1990}
Kaufman, L. and P.~J. Rousseeuw (1990).
\newblock Partitioning around medoids (program {PAM}).
\newblock In L.~Kaufman and P.~J. Rousseeuw (Eds.), {\em {Finding Groups in
  Data: An Introduction to Cluster Analysis}}, Chapter~2, pp.\  68--125. New
  York: John Wiley \& Sons.

\bibitem[\protect\citeauthoryear{Lazarsfeld and Henry}{Lazarsfeld and
  Henry}{1968}]{Lazarsfeld1968}
Lazarsfeld, P.~F. and N.~W. Henry (1968).
\newblock {\em {Latent Structure Analysis}}.
\newblock Boston: Houghton Mifflin.

\bibitem[\protect\citeauthoryear{Lesnard}{Lesnard}{2010}]{Lesnard2010}
Lesnard, L. (2010).
\newblock Setting cost in optimal matching to uncover contemporaneous
  socio-temporal patterns.
\newblock {\em Sociological Methods \& Research\/}~{\em 38\/}(3), 389--419.

\bibitem[\protect\citeauthoryear{Levenshtein}{Levenshtein}{1966}]{Levenshstein1966}
Levenshtein, V.~I. (1966).
\newblock Binary codes capable of correcting deletions, insertions, and
  reversals.
\newblock {\em Soviet Physics Doklady\/}~{\em 10\/}(8), 707--710.

\bibitem[\protect\citeauthoryear{Linzer and Lewis}{Linzer and
  Lewis}{2011}]{Linzer2011}
Linzer, D.~A. and J.~B. Lewis (2011).
\newblock {poLCA}: an \textsf{R} package for polytomous variable latent class
  analysis.
\newblock {\em Journal of Statistical Software\/}~{\em 42\/}(10), 1--29.

\bibitem[\protect\citeauthoryear{Mallows}{Mallows}{1957}]{Mallows1957}
Mallows, C.~L. (1957).
\newblock Non-null ranking models.
\newblock {\em Biometrika\/}~{\em 44\/}(1/2), 114--130.

\bibitem[\protect\citeauthoryear{McNicholas, Murphy, McDaid, and
  Frost}{McNicholas et~al.}{2010}]{McNicholas2010c}
McNicholas, P.~D., T.~B. Murphy, A.~F. McDaid, and D.~Frost (2010).
\newblock Serial and parallel implementations of model-based clustering via
  parsimonious {G}aussian mixture models.
\newblock {\em Computational Statistics \& Data Analysis\/}~{\em 54\/}(3),
  711--723.

\bibitem[\protect\citeauthoryear{McVicar}{McVicar}{2000}]{McVicar2000}
McVicar, D. (2000).
\newblock {Status 0 four years on: young people and social exclusion in
  Northern Ireland}.
\newblock {\em Labour Market Bulletin\/}~{\em 14}, 114--119.

\bibitem[\protect\citeauthoryear{McVicar and Anyadike-Danes}{McVicar and
  Anyadike-Danes}{2002}]{mvad2002}
McVicar, D. and M.~Anyadike-Danes (2002).
\newblock {P}redicting successful and unsuccessful transitions from school to
  work by using sequence methods.
\newblock {\em Journal of the Royal Statistical Society: Series A (Statistics
  in Society)\/}~{\em 165\/}(2), 317--334.

\bibitem[\protect\citeauthoryear{Melnykov}{Melnykov}{2016a}]{Melnykov2016}
Melnykov, V. (2016a).
\newblock {\GG{20160101}}{M}odel-based biclustering of clickstream data.
\newblock {\em Computational Statistics \& Data Analysis\/}~{\em 93\/}(C),
  31--45.

\bibitem[\protect\citeauthoryear{Melnykov}{Melnykov}{2016b}]{ClickClust2016}
Melnykov, V. (2016b).
\newblock {\GG{20160202}}{ClickClust}: an \textsf{R} package for model-based
  clustering of categorical sequences.
\newblock {\em Journal of Statistical Software\/}~{\em 74\/}(9), 1--34.

\bibitem[\protect\citeauthoryear{Menardi}{Menardi}{2011}]{Menardi2011}
Menardi, G. (2011).
\newblock Density-based silhouette diagnostics for clustering methods.
\newblock {\em Statistics and Computing\/}~{\em 21\/}(3), 295--308.

\bibitem[\protect\citeauthoryear{Meng and Rubin}{Meng and
  Rubin}{1993}]{Meng1993}
Meng, X.~L. and D.~R. Rubin (1993).
\newblock Maximum likelihood estimation via the {ECM} algorithm: a general
  framework.
\newblock {\em Biometrika\/}~{\em 80\/}(2), 267--278.

\bibitem[\protect\citeauthoryear{Mu{\~{n}}oz-Bull{\'{o}}n and
  Malo}{Mu{\~{n}}oz-Bull{\'{o}}n and Malo}{2003}]{MunozBullon2003}
Mu{\~{n}}oz-Bull{\'{o}}n, F. and M.~A. Malo (2003).
\newblock {Employment status mobility from a life-cycle perspective: a sequence
  analysis of work-histories in the BHPS}.
\newblock {\em Demographic Research\/}~{\em 9\/}(7), 119--162.

\bibitem[\protect\citeauthoryear{Murphy and Murphy}{Murphy and
  Murphy}{2020}]{Murphy2020}
Murphy, K. and T.~B. Murphy (2020).
\newblock {G}aussian parsimonious clustering models with covariates and a noise
  component.
\newblock {\em Advances in Data Analysis and Classification\/}~{\em 14\/}(2),
  293--325.

\bibitem[\protect\citeauthoryear{Murphy, Murphy, Piccarreta, and
  Gormley}{Murphy et~al.}{2021}]{MEDseqR2021}
Murphy, K., T.~B. Murphy, R.~Piccarreta, and I.~C. Gormley (2021).
\newblock {\em \texttt{\textup{MEDseq}}: mixtures of exponential-distance
  models with covariates}.
\newblock \textsf{R} package version 1.3.2.

\bibitem[\protect\citeauthoryear{Murphy and Martin}{Murphy and
  Martin}{2003}]{Murphy2003}
Murphy, T.~B. and D.~Martin (2003).
\newblock {M}ixtures of distance-based models for ranking data.
\newblock {\em Computational Statistics \& Data Analysis\/}~{\em 41\/}(3--4),
  645--655.

\bibitem[\protect\citeauthoryear{O'Hagan, Murphy, Scrucca, and Gormley}{O'Hagan
  et~al.}{2019}]{OHagan2019}
O'Hagan, A., T.~B. Murphy, L.~Scrucca, and I.~C. Gormley (2019).
\newblock Investigation of parameter uncertainty in clustering using a
  {G}aussian mixture model via jackknife, bootstrap and weighted likelihood
  bootstrap.
\newblock {\em Computational Statistics\/}~{\em 34\/}(4), 1779--1813.

\bibitem[\protect\citeauthoryear{Pamminger and
  Fr{\"{u}}hwirth-Schnatter}{Pamminger and
  Fr{\"{u}}hwirth-Schnatter}{2010}]{Pamminger2010}
Pamminger, C. and S.~Fr{\"{u}}hwirth-Schnatter (2010).
\newblock {M}odel-based clustering of categorical time series.
\newblock {\em Bayesian Analysis\/}~{\em 5\/}(2), 345--368.

\bibitem[\protect\citeauthoryear{Piccarreta and Studer}{Piccarreta and
  Studer}{2019}]{Piccarreta2019}
Piccarreta, R. and M.~Studer (2019).
\newblock Holistic analysis of the life course: methodological challenges and
  new perspectives.
\newblock {\em Advances in Life Course Research\/}~{\em 41}, 100251.

\bibitem[\protect\citeauthoryear{{\textsf{R} Core Team}}{{\textsf{R} Core
  Team}}{2021}]{R2021}
{\textsf{R} Core Team} (2021).
\newblock {\em \textsf{R}: a language and environment for statistical
  computing}.
\newblock Vienna, Austria: \textsf{R} Foundation for Statistical Computing.

\bibitem[\protect\citeauthoryear{Rousseeuw}{Rousseeuw}{1987}]{Rousseeuw1987}
Rousseeuw, P.~J. (1987).
\newblock Silhouettes: a graphical aid to the interpretation and validation of
  cluster analysis.
\newblock {\em Computational and Applied Mathematics\/}~{\em 20}, 53--65.

\bibitem[\protect\citeauthoryear{Schwarz}{Schwarz}{1978}]{Schwarz1978}
Schwarz, G. (1978).
\newblock {E}stimating the dimension of a model.
\newblock {\em The Annals of Statistics\/}~{\em 6\/}(2), 461--464.

\bibitem[\protect\citeauthoryear{Studer}{Studer}{2013}]{WeightedCluster2013}
Studer, M. (2013).
\newblock {W}eighted{C}luster library manual: a practical guide to creating
  typologies of trajectories in the social sciences with \textsf{R}.
\newblock Technical report, LIVES Working Papers 24.

\bibitem[\protect\citeauthoryear{Studer}{Studer}{2018}]{Studer2018}
Studer, M. (2018).
\newblock Divisive property-based and fuzzy clustering for sequence analysis.
\newblock In G.~Ritschard and M.~Studer (Eds.), {\em Sequence Analysis and
  Related Approaches: Innovative Methods and Applications}, pp.\  223--239.
  Cham: Springer International Publishing.

\bibitem[\protect\citeauthoryear{Studer and Ritschard}{Studer and
  Ritschard}{2016}]{Studer2016}
Studer, M. and G.~Ritschard (2016).
\newblock What matters in differences between life trajectories: a comparative
  review of sequence dissimilarity measures.
\newblock {\em Journal of the Royal Statistical Society: Series A (Statistics
  in Society)\/}~{\em 179\/}(2), 481--511.

\bibitem[\protect\citeauthoryear{Ward}{Ward}{1963}]{Ward1963}
Ward, Jr., J.~H. (1963).
\newblock Hierarchical grouping to optimize an objective function.
\newblock {\em Journal of the American Statistical Association\/}~{\em
  58\/}(301), 236--244.

\bibitem[\protect\citeauthoryear{Wu}{Wu}{2000}]{Wu2000}
Wu, L.~L. (2000).
\newblock Some comments on sequence analysis and optimal matching methods in
  sociology: review and prospect.
\newblock {\em Sociological Methods \& Research\/}~{\em 29\/}(1), 41--64.

\bibitem[\protect\citeauthoryear{Xu, Chen, and Mantell}{Xu
  et~al.}{2013}]{Xu2013}
Xu, C., J.~Chen, and H.~Mantell (2013).
\newblock Pseudo-likelihood-based {B}ayesian information criterion for variable
  selection in survey data.
\newblock {\em Survey Methodology\/}~{\em 39\/}(2), 303--322.

\end{thebibliography}

\let\appendixpagenameorig\appendixpagename
\label{key}\renewcommand{\appendixpagename}{\Large\appendixpagenameorig}
\makeatletter
\renewcommand\section{\@startsection{section}{1}{\z@}%
	{-3.5ex \@plus -1ex \@minus -.2ex}%
	{2.5ex \@plus.2ex}%
	{\normalfont\large\bfseries}}
\makeatother
\normalsize
\clearpage

\begin{appendices}
	\label{Section:Appendices}
	\setcounter{table}{0}
	\setcounter{figure}{0}
	\renewcommand\thetable{\thesection.\arabic{table}}
	\renewcommand\thefigure{\thesection.\arabic{figure}}
	
	\section[The MEDseq Model Family: Parameter Counts]{The MEDseq Model Family: Parameter Counts}
	\label{Section:ParameterCount}
	
	The models in the MEDseq family differ only in their treatment of the precision parameters, which differentiate the Hamming distance and the weighted variants thereof. The BIC is used in order to choose between the $8$ model types, identify the optimal $G$, and guide the inclusion of gating covariates. Table \ref{Table:params} summarises the number of free parameters $k$ in the BIC penalty term under each MEDseq model type, in order to demonstrate the increasing level of complexity in moving from the most parsimonious \textsf{CCN} model to the most heavily parameterised \textsf{UU} model. 
	
	The number of parameters contributing to each $\widehat{\boldsymbol{\theta}}_g$ estimate notably depends on the number of states represented across all cases in each time point. Note also that parameters relating to $\smash{\widehat{\theta}_{g, t}}$ corresponding to \emph{estimated} precision parameters are counted, while those associated with fixed precision parameter values of $0$ are not counted. Similarly, precision parameters estimated as $0$ are counted, but precision parameters fixed at $0$ associated with the noise component are not. 
	
	The number of gating network parameters is not accounted for in Table \ref{Table:params}. When covariates are included, there are $\left(r+1\right)\times \left(G-1\right)$ or $\left(r+1\right)\times \left(G-2\right) + 1$ extra parameters --- under the GN and NGN settings, respectively --- where $r+1$~is the dimension of the associated design matrix, including the intercept term. When $\boldsymbol{\tau}$ is not covariate-dependent, there are $G-1$ extra parameters when $\boldsymbol{\tau}$ is unconstrained or only $1$ extra parameter if $\boldsymbol{\tau}$ is constrained and the model includes a noise component, in which case $\tau_0$ is allowed to vary.
	\begin{table}[H]
		\caption{Number of estimated parameters under each MEDseq model type. Models with names ending with the letter \textsf{N}, indicating the presence of a noise component for which the single precision parameter is fixed to $0$, behave like the corresponding model without this component for all other components. Thus, $\lambda$ and all subscript variants thereof refer here to the non-noise components only.\label{Table:params}}
		\centering
		\scriptsize
		\extrarowheight 2.5pt
		
		\begin{tabular}[pos=center]{c c | c c | c c}
			\specialrule{.1em}{.01em}{.01em} 
			\multirow{2}{*}{Model} & \multirow{2}{*}{Precision} & \multirow{2}{*}{$\lambda_g$~(Clusters)} & \multirow{2}{*}{$\lambda_t$~(Time Points)} & \multicolumn{2}{c}{Number of Parameters}\\
			\cline{5-6}
			& & &  & Central Sequence(s) & Precision\\ 
			\hline\hline
			\textsf{CC} & \multirow{2}{*}{$\lambda_{g,t} = \lambda$} & \multirow{2}{*}{Constrained} & \multirow{2}{*}{Constrained} & $G\sum_{t=1}^T\left(v_t - 1\right)$ & $1$\\
			\textsf{CCN} &  &  &  & $\left(G-1\right)\sum_{t=1}^T\left(v_t - 1\right)$ & $\indicator{G > 1}$\\
			\hline
			\textsf{UC} & \multirow{2}{*}{$\lambda_{g,t} = \lambda_g$} & \multirow{2}{*}{Unconstrained} & \multirow{2}{*}{Constrained} & $G\sum_{t=1}^T\left(v_t - 1\right)$ & $G$\\
			\textsf{UCN} &  &  &  & $\left(G-1\right)\sum_{t=1}^T\left(v_t - 1\right)$ &  $G-1$\\
			\hline
			\textsf{CU} & \multirow{2}{*}{$\lambda_{g,t} = \lambda_t$} & \multirow{2}{*}{Constrained} & \multirow{2}{*}{Unconstrained} & $G\sum_{t=1}^T\left(v_t - 1\right)$ &  $T$\\
			\textsf{CUN} &  &  &  & $\left(G-1\right)\sum_{t=1}^T\left(v_t - 1\right)$ &  $\indicator{G > 1}T$\\
			\hline
			\textsf{UU} & \multirow{2}{*}{$\lambda_{g,t} = \lambda_{g,t}$} & \multirow{2}{*}{Unconstrained} & \multirow{2}{*}{Unconstrained} & $G\sum_{t=1}^T\left(v_t - 1\right)$ & $GT$\\
			\textsf{UUN} & & & & $\left(G-1\right)\sum_{t=1}^T\left(v_t - 1\right)$ & $\left(G - 1\right)T$ \\
			\specialrule{.1em}{.01em}{.01em} 
		\end{tabular}
	\end{table}
	
	\section[Estimating MEDseq Precision Parameters]{Estimating MEDseq Precision Parameters}
	\label{Section:AllSteps}
	\setcounter{table}{0}
	\setcounter{figure}{0}
	\renewcommand\thetable{\thesection.\arabic{table}}
	\renewcommand\thefigure{\thesection.\arabic{figure}}
	
	For fixed $\boldsymbol{\theta}$, the PMF in \eqref{eq:EDmodel} belongs to the exponential family with natural parameter $\lambda$. Thus, under any distance metric, the method of moments estimate of $\lambda$ is equal to the~MLE. Hence, with $\smash{\widehat{\boldsymbol{\theta}}}$ already estimated as per Section \ref{Section:theta}, $\smash{\widehat{\lambda}}$ ensures that the expected distance of observations from $\smash{\widehat{\boldsymbol{\theta}}}$ is equal to the observed average distance from $\smash{\widehat{\boldsymbol{\theta}}}$, since the solution~of
	\[\frac{\partial\ell\negthinspace\left(\lambda\given\mathbf{S},\smash{\widehat{\boldsymbol{\theta}}},\mathrm{d}\right)}{n\partial\lambda} = \frac{\sum_{\boldsymbol{\sigma}\,\in\,\bm{\mathcal{S}}_v^T}\mathrm{d}\big(\boldsymbol{\sigma},\widehat{\boldsymbol{\theta}}\big)\exp\negthinspace\big(\negthinspace-\negthinspace\lambda\mathrm{d}\big(\boldsymbol{\sigma},\widehat{\boldsymbol{\theta}}\big)\big)}{\sum_{\boldsymbol{\sigma}\,\in\,\bm{\mathcal{S}}_v^T}\exp\negthinspace\big(\negthinspace-\negthinspace\lambda\mathrm{d}\big(\boldsymbol{\sigma},\widehat{\boldsymbol{\theta}}\big)\big)} - \frac{1}{n}\sum_{i=1}^n \mathrm{d}\negthinspace\left(\mathbf{s}_i,\smash{\widehat{\boldsymbol{\theta}}}\right)\]
	implies
	\begin{equation}
	\mathbb{E}_\lambda\negthinspace\left(\mathrm{d}\negthinspace\left(\mathbf{S},\smash{\widehat{\boldsymbol{\theta}}}\right)\right) = \frac{\sum_{\boldsymbol{\sigma}\,\in\,\bm{\mathcal{S}}_v^T}\mathrm{d}\big(\boldsymbol{\sigma},\widehat{\boldsymbol{\theta}}\big)\exp\negthinspace\big(\negthinspace-\negthinspace\lambda\mathrm{d}\big(\boldsymbol{\sigma},\widehat{\boldsymbol{\theta}}\big)\big)}{\sum_{\boldsymbol{\sigma}\,\in\,\bm{\mathcal{S}}_v^T}\exp\negthinspace\big(\negthinspace-\negthinspace\lambda\mathrm{d}\big(\boldsymbol{\sigma},\widehat{\boldsymbol{\theta}}\big)\big)}= \overline{\mathrm{d}}\big(\mathbf{S},\widehat{\boldsymbol{\theta}}\big) =\frac{1}{n}\sum_{i=1}^n\mathrm{d}\big(\mathbf{s}_i,\widehat{\boldsymbol{\theta}}\big).\label{eq:MOMmle}
	\end{equation}
	\indent Under the Hamming distance, the value of the expectation in \eqref{eq:MOMmle} holds for any arbitrary reference sequence in place of $\widehat{\boldsymbol{\theta}}$. As the denominator in \eqref{eq:MOMmle} --- corresponding to the normalising constant in \eqref{eq:PsiH}, under the Hamming distance --- is a function of $\lambda$, it is crucial that it exists in closed form in order to estimate the precision parameter. Hence, with known $\smash{\widehat{\boldsymbol{\theta}}}$, the MLE for $\lambda$ for an unweighted single-component \textsf{CC} model can be obtained as follows:
	\begin{align*}
	\ell\big(\lambda\given\mathbf{S},\widehat{\boldsymbol{\theta}},\mathrm{d_H}\big) &= -\lambda n \overline{\mathrm{d}}_{\mathrm{H}}\big(\mathbf{S},\widehat{\boldsymbol{\theta}}\big) - nT\log\negthinspace\left(\left(v-1\right)e^{-\lambda} + 1\right),\\
	\frac{\partial\ell\left(\cdot\right)}{\partial\lambda} &= \frac{nT\left(v-1\right)}{e^\lambda + \left(v-1\right)} - n\overline{\mathrm{d}}_{\mathrm{H}}\big(\mathbf{S},\widehat{\boldsymbol{\theta}}\big),\\
	\therefore\:\widehat{\lambda} & =  \log\Bigg(\negthinspace\left(v-1\right)\bigg(\frac{T}{\overline{\mathrm{d}}_{\mathrm{H}}\big(\mathbf{S},\widehat{\boldsymbol{\theta}}\big)} - 1\bigg)\Bigg),
	\end{align*}
	which notably relies on the inverse of the average Hamming distance normalised by the sequence length $T$. However, this can yield a negative value for $\smash{\widehat{\lambda}}$. Recall that we only consider $\lambda \geq 0$. Since all distances are non-negative and typically not identical,  $\frac{\partial\ell\left(\cdot\right)}{\partial\lambda}$ is negative $\forall\:\lambda>0$ in the case where the sufficient statistic $\smash{\overline{\mathrm{d}}_{\mathrm{H}}\big(\mathbf{S},\widehat{\boldsymbol{\theta}}\big) > v^{-1}T\left(v-1\right)}$, with $\lim_{\lambda\to\infty}\negthinspace\frac{\partial\ell\left(\cdot\right)}{\partial\lambda} = -n\overline{\mathrm{d}}_{\mathrm{H}}\big(\mathbf{S},\widehat{\boldsymbol{\theta}}\big)$. Thus,
	\[\widehat{\lambda}=\max\negthinspace\Bigg(0, \log\negthinspace\bigg(\negthinspace\left(v-1\right)\Big(\frac{T}{\overline{\mathrm{d}}_{\mathrm{H}}\big(\mathbf{S}, \widehat{\boldsymbol{\theta}}\big)} - 1\Big)\bigg)\Bigg).\] 
	\noindent When $\smash{\overline{\mathrm{d}}_{\mathrm{H}}\big(\mathbf{S},\widehat{\boldsymbol{\theta}}\big) < v^{-1}T\left(v-1\right)}$, such that $\widehat{\lambda} > 0$, the identity $\log\negthinspace\left(c\left(a/b - 1\right)\right) = \log\negthinspace\left(c\right) + \log\negthinspace\left(a-b\right) - \log\negthinspace\left(b\right)$ is used for numerical stability, otherwise $\widehat{\lambda}$ is set to $0$. When sampling weights are included, following the same steps as above yields the corresponding estimate
	\begin{equation}
	\widehat{\lambda}=\max\negthinspace\left(0, \log\negthinspace\left(v-1\right) + \log\negthinspace\bigg(\frac{Tn}{\sum_{i=1}^nw_i\mathrm{d_H}\big(\mathbf{s}_i,\widehat{\boldsymbol{\theta}}\big)} - 1\bigg)\right).\label{eq:lambdaG1CC}
	\end{equation} 
	\indent The ECM algorithm is employed when $G>1$, in which case the CM-step for $\smash{\widehat{\lambda}^{\left(m+1\right)}}$ under a \textsf{CC} MEDseq mixture model with sampling weights is given by
	\begin{align}
	\frac{\partial\ell^{\mathbf{w}}_c\left(\cdot\right)}{\partial\lambda} &= \frac{T\left(v-1\right)\sum_{i=1}^n\sum_{g=1}^Gz_{i,g}w_i}{e^\lambda + \left(v-1\right)} -\sum_{i=1}^{n}\sum_{g=1}^{G}z_{i,g}w_i\mathrm{d_H}\big(\mathbf{s}_i,\widehat{\boldsymbol{\theta}}_g\big),\nonumber\\
	\therefore\:\widehat{\lambda}^{\left(m+1\right)} &= \max\negthinspace\left(0,\log\negthinspace\left(v-1\right) + \log\negthinspace\Bigg(\frac{Tn}{\sum_{i=1}^n\sum_{g=1}^{G}\widehat{z}_{i,g}^{\left(m+1\right)}w_i\mathrm{d_H}\big(\mathbf{s}_i,\widehat{\boldsymbol{\theta}}_g^{\left(m+1\right)}\big)} - 1\Bigg)\right).\label{eq:lambdaGgCC}
	\end{align}
	\noindent As per \eqref{eq:lambdaG1CC}, this requires the current estimate of each component's central sequence. When there are no sampling weights, one need only drop the $w_i$ terms from \eqref{eq:lambdaG1CC} and \eqref{eq:lambdaGgCC} to estimate the precision parameters of unweighted MEDseq~\mbox{models}. While $\smash{\widehat{\lambda}}$ can potentially be estimated as zero, the inclusion of a noise component~in the \textsf{CCN}, \textsf{UCN}, \textsf{CUN}, and \textsf{UUN} models makes this explicit, by restricting one cluster to have $\smash{\lambda_{g,t}=0\:\forall\:t=1,\ldots,T}$. 
	
	However, when $\smash{\widehat{\lambda}_{g,t}}$ is estimated as zero rather than fixed to zero, the corresponding $\smash{\theta_{g,t}}$ parameter must be estimated, as it affects the likelihood indirectly through its role in estimating the precision parameter(s). In particular --- taking the \textsf{UU} model as an example --- all state values in the $t$-th sequence position with non-zero $\widehat{z}_{i, g}^{\left(m+1\right)}$ are identical to $\widehat{\theta}_{g, t}^{\left(m+1\right)}$ when the corresponding denominator in Table \ref{Table:PrecisionMsteps} evaluates to zero, such that $\smash{\widehat{\lambda}_{g, t}^{\left(m+1\right)} \to \infty}$.
	
	Expressions for the weighted complete data pseudo likelihood functions for all model types in the MEDseq family are given in Table \ref{Table:WeightedCompleteLike}. All models are written here as though gating network covariates $\mathbf{x}_i$ are included. However, the gating networks of models with a noise component are written in the NGN form employed by the optimal model identified in Section \ref{Section:MVADresults} rather than the GN form, i.e. it is assumed that $\tau_0$ is constant, meaning the covariates do not affect the probability of belonging to the noise component (see Section~\ref{Section:Gating}).
	
	Table \ref{Table:PrecisionMsteps} outlines the corresponding CM-steps for the precision parameter(s). All derivations closely follow the same steps as in \eqref{eq:lambdaGgCC} for the \textsf{CC} model and the normalised sampling weights are accounted for in all cases. These formulas can be simplified somewhat for unweighted models and/or models without gating covariates. Recall that the first letter of the model~name denotes whether the precision parameters are constrained/unconstrained across clusters, the second denotes the same across time points (i.e. sequence positions), and model names ending with the letter \textsf{N} include a noise component. 
	\begin{table}[H]
		\caption{Weighted complete data pseudo likelihood functions for all MEDseq model types, which differ according to the constraints imposed on the precision parameters across clusters and/or time points. The expressions for the various weighted Hamming distance metric variants employed, and the associated normalising constants, are given in full. \label{Table:WeightedCompleteLike}}
		\centering
		\fontsize{11}{12}\selectfont
			\begin{tabular}[pos=center]{c | l}
				\specialrule{.1em}{.01em}{.01em} 
				Model & Weighted Complete Data Pseudo Likelihood\\
				\hline\hline
				\\
				\textsf{CC}& $\prod_{i=1}^n\left\lbrack\prod_{g=1}^G \left(\tau_g\negthinspace\left(\mathbf{x}_i\right)\frac{\exp\left(-\lambda\sum_{t=1}^T\indicator{s_{i,t} \neq \theta_{g,t}}\right)}{\left(\left(v-1\right)e^{-\lambda} + 1\right)^T}\right)^{z_{i,g}}\right\rbrack^{w_i}$\\[5ex]
				\textsf{UC}& $\prod_{i=1}^n\left\lbrack\prod_{g=1}^G \left(\tau_g\negthinspace\left(\mathbf{x}_i\right)\frac{\exp\left(-\lambda_g\negthinspace\sum_{t=1}^T\indicator{s_{i,t} \neq \theta_{g,t}}\right)}{\left(\left(v-1\right)e^{-\lambda_g} + 1\right)^T}\right)^{z_{i,g}}\right\rbrack^{w_i}$\\[5ex]
				\textsf{CU}& $\prod_{i=1}^n\left\lbrack\prod_{g=1}^G \left(\tau_g\negthinspace\left(\mathbf{x}_i\right)\frac{\exp\left(-\sum_{t=1}^T\lambda_t\indicator{s_{i,t} \neq \theta_{g,t}}\right)}{\prod_{t=1}^T\left(\left(v-1\right)e^{-\lambda_t} + 1\right)}\right)^{z_{i,g}}\right\rbrack^{w_i}$\\[5ex]
				\textsf{UU}& $\prod_{i=1}^n\left\lbrack\prod_{g=1}^G \left(\tau_g\negthinspace\left(\mathbf{x}_i\right)\frac{\exp\left(-\sum_{t=1}^T\lambda_{gt}\indicator{s_{i,t} \neq \theta_{g,t}}\right)}{\prod_{t=1}^T\left(\left(v-1\right)e^{-\lambda_{g,t}} + 1\right)}\right)^{z_{i,g}}\right\rbrack^{w_i}$\\[5ex]
				\textsf{CCN}& $\prod_{i=1}^n\left\lbrack\left(\prod_{g=1}^{G-1}\left( \tau_g\negthinspace\left(\mathbf{x}_i\right)\frac{\exp\left(-\lambda\sum_{t=1}^T\indicator{s_{i,t} \neq \theta_{g,t}}\right)}{\left(\left(v-1\right)e^{-\lambda} + 1\right)^T}\right)^{z_{i,g}}\right)\left(\frac{\tau_0}{v^T}\right)^{z_{i,0}}\right\rbrack^{w_i}$\\[5ex]
				\textsf{UCN}&$\prod_{i=1}^n\left\lbrack\left(\prod_{g=1}^{G-1}\left( \tau_g\negthinspace\left(\mathbf{x}_i\right)\frac{\exp\left(-\lambda_g\negthinspace\sum_{t=1}^T\indicator{s_{i,t} \neq \theta_{g,t}}\right)}{\left(\left(v-1\right)e^{-\lambda_g} + 1\right)^T}\right)^{z_{i,g}}\right)\left(\frac{\tau_0}{v^T}\right)^{z_{i,0}}\right\rbrack^{w_i}$\\[5ex]
				\textsf{CUN}& $\prod_{i=1}^n\left\lbrack\left(\prod_{g=1}^{G-1}\left( \tau_g\negthinspace\left(\mathbf{x}_i\right)\frac{\exp\left(-\sum_{t=1}^T\lambda_t\indicator{s_{i,t} \neq \theta_{g,t}}\right)}{\prod_{t=1}^T\left(\left(v-1\right)e^{-\lambda_t} + 1\right)}\right)^{z_{i,g}}\right)\left(\frac{\tau_0}{v^T}\right)^{z_{i,0}}\right\rbrack^{w_i}$\\[5ex]
				\textsf{UUN}&$\prod_{i=1}^n\left\lbrack\left(\prod_{g=1}^{G-1}\left( \tau_g\negthinspace\left(\mathbf{x}_i\right)\frac{\exp\left(-\sum_{t=1}^T\lambda_{g,t}\indicator{s_{i,t} \neq \theta_{g,t}}\right)}{\prod_{t=1}^T\left(\left(v-1\right)e^{-\lambda_{g,t}} + 1\right)}\right)^{z_{i,g}}\right)\left(\frac{\tau_0}{v^T}\right)^{z_{i,0}}\right\rbrack^{w_i}$\\[4.25ex]
				\specialrule{.1em}{.01em}{.01em} 
			\end{tabular}
	\end{table}
	\begin{table}[H]
		\caption{CM-steps for the precision parameter(s) of the non-noise components for all MEDseq model types, which differ according to the constraints imposed across clusters and/or time points. \label{Table:PrecisionMsteps}}
		\centering
		\fontsize{11}{12}\selectfont
		\begin{tabular}[pos=center]{c | l}
			\specialrule{.1em}{.01em}{.01em} 
			Model & Precision Parameter CM-steps\\
			\hline\hline
			\\
			\textsf{CC}&  $\widehat{\lambda}^{\left(m+1\right)}=\max\negthinspace\left(0,\log\negthinspace\left(v-1\right) + \log\negthinspace\bigg(\frac{Tn}{\sum_{i=1}^n\sum_{g=1}^G\widehat{z}_{i,g}^{\left(m+1\right)}w_i\mathrm{d_H}\negthinspace\left(\mathbf{s}_i,\widehat{\boldsymbol{\theta}}_g^{\left(m+1\right)}\right)} - 1 \bigg)\right)$\\[5ex]
			\textsf{UC}& $\widehat{\lambda}_g^{\left(m+1\right)}=\max\negthinspace\left(0,\log\negthinspace\left(v-1\right) + \log\negthinspace\bigg(\frac{T\sum_{i=1}^n\widehat{z}_{i,g}^{\left(m+1\right)}w_i}{\sum_{i=1}^n\widehat{z}_{i,g}^{\left(m+1\right)}w_i\mathrm{d_H}\negthinspace\left(\mathbf{s}_i,\widehat{\boldsymbol{\theta}}_g^{\left(m+1\right)}\right)} - 1 \bigg)\right)$\\[5ex]
			\textsf{CU}& $\widehat{\lambda}_t^{\left(m+1\right)}=\max\negthinspace\left(0,\log\negthinspace\left(v-1\right) + \log\negthinspace\bigg(\frac{n}{\sum_{i=1}^n\sum_{g=1}^G\widehat{z}_{i,g}^{\left(m+1\right)}w_i\indicator{s_{i,t} \neq \widehat{\theta}_{g,t}^{\left(m+1\right)}}} - 1 \bigg)\right)$\\[5ex]
			\textsf{UU}& $\widehat{\lambda}_{g,t}^{\left(m+1\right)}=\max\negthinspace\left(0,\log\negthinspace\left(v-1\right) + \log\negthinspace\bigg(\frac{\sum_{i=1}^n\widehat{z}_{i,g}^{\left(m+1\right)}w_i}{\sum_{i=1}^n\widehat{z}_{i,g}^{\left(m+1\right)}w_i\indicator{s_{i,t} \neq \widehat{\theta}_{g,t}^{\left(m+1\right)}}} - 1 \bigg)\right)$\\[5ex]
			\textsf{CCN}& $\widehat{\lambda}^{\left(m+1\right)}=\max\negthinspace\left(0,\log\negthinspace\left(v-1\right) + \log\negthinspace\bigg(\frac{T\sum_{i=1}^n\sum_{g=1}^{G-1}\widehat{z}_{i,g}^{\left(m+1\right)}w_i}{\sum_{i=1}^n\sum_{g=1}^{G-1}\widehat{z}_{i,g}^{\left(m+1\right)}w_i\mathrm{d_H}\negthinspace\left(\mathbf{s}_i,\widehat{\boldsymbol{\theta}}_g^{\left(m+1\right)}\right)} - 1 \bigg)\right)$\\[5ex]
			\textsf{UCN}& $\widehat{\lambda}_g^{\left(m+1\right)}=\max\negthinspace\left(0,\log\negthinspace\left(v-1\right) + \log\negthinspace\bigg(\frac{T\sum_{i=1}^n\widehat{z}_{i,g}^{\left(m+1\right)}w_i}{\sum_{i=1}^n\widehat{z}_{i,g}^{\left(m+1\right)}w_i\mathrm{d_H}\negthinspace\left(\mathbf{s}_i,\widehat{\boldsymbol{\theta}}_g^{\left(m+1\right)}\right)} - 1 \bigg)\right)$\\[5ex]
			\textsf{CUN}&  $\widehat{\lambda}_t^{\left(m+1\right)}=\max\negthinspace\left(0,\log\negthinspace\left(v-1\right) + \log\negthinspace\bigg(\frac{\sum_{i=1}^n\sum_{g=1}^{G-1}\widehat{z}_{i,g}^{\left(m+1\right)}w_i}{\sum_{i=1}^n\sum_{g=1}^{G-1}\widehat{z}_{i,g}^{\left(m+1\right)}w_i\indicator{s_{i,t} \neq \widehat{\theta}_{g,t}^{\left(m+1\right)}}} - 1 \bigg)\right)$\\[5ex]
			\textsf{UUN}& $\widehat{\lambda}_{g,t}^{\left(m+1\right)}=\max\negthinspace\left(0,\log\negthinspace\left(v-1\right) + \log\negthinspace\bigg(\frac{\sum_{i=1}^n\widehat{z}_{i,g}^{\left(m+1\right)}w_i}{\sum_{i=1}^n\widehat{z}_{i,g}^{\left(m+1\right)}w_i\indicator{s_{i,t} \neq \widehat{\theta}_{g,t}^{\left(m+1\right)}}} - 1 \bigg)\right)$\\[4.25ex]
			\specialrule{.1em}{.01em}{.01em} 
		\end{tabular}
	\end{table}

	\section[MVAD Data: Gating Network Coefficients]{MVAD Data: Gating Network Coefficients}
	\label{Section:AllGating}
	\setcounter{table}{0}
	\setcounter{figure}{0}
	\renewcommand\thetable{\thesection.\arabic{table}}
	\renewcommand\thefigure{\thesection.\arabic{figure}}
	
	Multinomial logistic regression coefficients and associated WLBS standard errors of the NGN gating network of a $G=11$ \textsf{UUN} model with the GCSE5eq covariate are provided~in Table \ref{Table:mvadGating}. For completeness, coefficients and standard errors for an otherwise equivalent model with all covariates included (except those used to define the sampling weights) are given in Table \ref{Table:mvadAllGating}. Such a model achieves a BIC value of $-93111.30$ (see Table \ref{Table:mvadBackwardStep}), compared~to $-92953.85$ for the optimal model with only the GCSE5eq covariate detailed in Section \ref{Section:MVADresults}. 
	
	Notably, $G=11$ and the \textsf{UUN} model type are both also optimal for the model with all covariates included. Furthermore, both models yield identical SPS labels, derived from $\smash{\widehat{\boldsymbol{\theta}}_g}$. Thus, Table \ref{Table:mvadGating} and Table \ref{Table:mvadAllGating} share the same reference level. In the latter case, we caution that the covariates `Livboth' and `Funemp' were measured after the observation period had begun, which complicates interpretation. In particular, subjects' living arrangements were recorded after two years, and the father's employment status was recorded in the final month of June 1999.
	\begin{table}[H]
		\caption{Multinomial logistic regression coefficients and associated WLBS standard errors (in parentheses) for the NGN gating network of the $G=11$ \textsf{UUN} model with all covariates included.\label{Table:mvadAllGating}}
		\centering
		\notsotiny
		\setlength{\tabcolsep}{2.5pt}
		\extrarowheight 2.5pt
		
		\begin{tabular}[pos=center]{c l | r r r r r r r r r r r r r r}
			\specialrule{.1em}{.01em}{.01em} 
			\multicolumn{2}{c|}{Cluster:\,$g$~(SPS)}  & \multicolumn{2}{c}{(Intercept)} & \multicolumn{2}{c}{Catholic} & \multicolumn{2}{c}{FMPR} & \multicolumn{2}{c}{Funemp} &  \multicolumn{2}{c}{GCSE5eq} & \multicolumn{2}{c}{Gender} & \multicolumn{2}{c}{Livboth}\\
			\hline\hline
			$1$&(SC,25)-(HE,45)           & \multicolumn{2}{c}{---}&\multicolumn{2}{c}{---}&\multicolumn{2}{c}{---}&\multicolumn{2}{c}{---}&\multicolumn{2}{c}{---}&\multicolumn{2}{c}{---}&\multicolumn{2}{c}{---}\\
			$2$&(FE,25)-(HE,45)               &$-0.97$&    $(0.60)$& $-0.65$&    $(0.41)$& $-0.49$&   $(0.42)$& $0.15$&   $(0.76)$&$-0.44$&    $(0.50)$&$-0.28$&   $(0.41)$& $0.73$&     $(0.45)$\\
			$3$&(SC,24)-(FE,36)-(HE,10)       &$-0.70$&    $(1.09)$& $-0.14$&    $(0.79)$& $-0.03$&   $(0.70)$&$-1.23$&   $(1.62)$&$-1.42$&    $(0.72)$&$-0.56$&   $(0.68)$& $0.95$&     $(0.78)$\\
			$4$&(SC,25)-(EM,45)               & $1.43$&    $(0.70)$& $-0.54$&    $(0.54)$& $-0.50$&   $(0.61)$& $0.94$&   $(0.95)$&$-2.15$&    $(0.56)$&$-0.71$&   $(0.56)$&$-0.45$&     $(0.56)$\\
			$5$&(TR,37)-(EM,33)               & $0.78$&    $(0.60)$& $-0.51$&    $(0.46)$& $-0.98$&   $(0.56)$& $0.48$&   $(0.73)$&$-3.24$&    $(0.54)$& $1.63$&   $(0.53)$&$-0.70$&     $(0.50)$\\
			$6$&(TR,22)-(EM,48)               & $1.27$&    $(0.67)$& $-0.50$&    $(0.46)$& $-0.64$&   $(0.48)$& $0.39$&   $(0.78)$&$-3.60$&    $(0.54)$& $0.63$&   $(0.46)$&$-0.12$&     $(0.49)$\\
			$7$&(TR,5)-(EM,65)                & $2.48$&    $(0.53)$& $-1.10$&    $(0.39)$& $-1.39$&   $(0.50)$& $0.79$&   $(0.71)$&$-3.89$&    $(0.49)$& $0.33$&   $(0.38)$&$-0.48$&     $(0.42)$\\
			$8$&(FE,22)-(EM,48)               & $1.24$&    $(0.61)$& $-0.06$&    $(0.41)$& $-0.61$&   $(0.44)$& $0.46$&   $(0.71)$&$-2.15$&    $(0.46)$&$-0.21$&   $(0.40)$&$-0.64$&     $(0.42)$\\
			$9$&(SC,10)-(FE,36)-(EM,24)       & $1.35$&    $(0.71)$& $-0.81$&    $(0.60)$& $-0.06$&   $(0.68)$& $1.43$&   $(0.94)$&$-3.10$&    $(0.57)$&$-0.11$&   $(0.54)$&$-0.35$&     $(0.52)$\\
			$10$&(TR,10)-(JL,2)-(TR,3)-(JL,55)& $0.65$&    $(1.54)$&  $0.13$&    $(0.88)$& $-0.47$&   $(0.61)$& $1.86$&   $(0.91)$&$-3.68$&    $(0.89)$&$-0.15$&   $(0.70)$&$-0.11$&     $(0.70)$\\
			\specialrule{.1em}{.01em}{.01em} 
		\end{tabular}
	\end{table}
	The MLR coefficients in Table \ref{Table:mvadGating} are estimated under a one-step approach, under which clustering and the relation of clusters to covariates are performed simultaneously. There are subtle interpretational differences between covariates serving as predictors of cluster-membership under a one-step approach compared to covariates enabling interpretation of~the type of observation characterising each cluster under a two-step approach. Fortunately, the MEDseq framework permits both types of analysis. Thus, we present coefficients obtained under two two-step approaches in Table \ref{Table:mvadTwoStep} in order to contrast them to those in Table \ref{Table:mvadGating}. 
	
	Firstly, a model without any covariates which is otherwise exactly equivalent to the\linebreak optimal one including only GCSE5eq as a covariate is fitted (BIC$=-93190.08$). Thereafter, MLR is used (with the GCSE5eq covariate, with the same reference level) firstly with the `soft' $\widehat{\mathbf{Z}}$ matrix as the response and secondly with the hard MAP partition used as the response. The sampling weights are accounted for in both cases; so too is the removal of the noise component, as per the NGN gating network setting shown in Table \ref{Table:mvadGating}, by appropriately renormalising $\widehat{\mathbf{Z}}$ prior to estimating each MLR (i.e. prior to constructing the MAP partition under the hard approach). Notably, there is little difference between the coefficients obtained under the one-step and soft two-step approaches. However, the estimates differ more greatly under the hard two-step approach, which suggests that relating covariates to hard partitions is inappropriate when the clusters are insufficiently homogeneous, as the hard partitions do not necessarily lead to appropriate responses for the MLR.
	\begin{table}[H]
		\caption{MLR coefficients obtained under two alternative two-step estimation approaches for the optimal model for the MVAD data. The first and second pair of rows give the intercept and slope coefficients when the soft cluster membership probabilities and hard assignments are used as the response, respectively.\label{Table:mvadTwoStep}}
		\centering
		\extrarowheight 2.5pt
		\scriptsize
		\setlength{\tabcolsep}{5pt}
		
		\begin{tabular}[pos=center]{c c | c c c c c c c c c c}
				\specialrule{.1em}{.01em}{.01em} 
				\multicolumn{2}{c|}{Cluster:\,$g$} & $1$ & $2$ & $3$ & $4$ & $5$ & $6$ & $7$ & $8$ & $9$ & $10$\\
				\hline\hline
				\multirow{2}{*}{Responses: $\widehat{\mathbf{z}}_i$} & (Intercept)&---&$-0.96$&$-0.38$&$0.58$&$1.02$&$1.18$&$1.70$&$0.60$&$0.92$&$0.90$\\
				&GCSE5eq&---&$-0.46$&$-1.31$&$-2.17$&$-3.43$&$-3.72$&$-4.06$&$-2.20$&$-3.17$&$-3.76$\\
				\specialrule{.1em}{.01em}{.01em} 
				\multirow{2}{*}{Responses: $\textrm{MAP}\negthinspace\left(\mathbf{\widehat{z}}_i\right)$} & (Intercept)&---&$-1.01$&$-0.47$&$0.57$&$0.95$&$1.17$&$1.69$&$0.60$&$0.91$&$0.86$\\
				&GCSE5eq&---&$-0.45$&$-1.23$&$-2.17$&$-3.65$&$-3.71$&$-4.30$&$-2.20$&$-3.23$&$-4.11$\\
				\specialrule{.1em}{.01em}{.01em} 
		\end{tabular}
	\end{table}

	Notably, similar conclusions are drawn when the coefficients in Table \ref{Table:mvadAllGating} are compared to those obtained (but not shown here) when similar soft and hard two-step approaches are used with the same set of `all' covariates; namely, the coefficients differ little between the one-step and soft two-step approach, but differ greatly between the one-step and hard two-step approach. However, the coefficients differ only in magnitude and not in sign in all pairwise comparisons.
\end{appendices}
\end{document}